\pdfoutput=1
\documentclass[usenatbib]{mn2e}
\usepackage{hyperref}
\usepackage{amsfonts,amsmath,amssymb}
\DeclareMathOperator{\sgn}{sgn}
\usepackage{txfonts}
\usepackage[]{graphicx}
\usepackage{bm}
\usepackage{ctable}
\usepackage{url}
\usepackage{breakurl}
\usepackage{fixltx2e} 
\usepackage[noabbrev]{cleveref}
\UseRawInputEncoding
\crefname{section}{\S\!}{\S\S\!}
\crefname{appendix}{App.}{Apps.}
\crefname{equation}{Eq.}{Eqs.}
\Crefname{equation}{Equation}{Equations}
\crefname{figure}{Fig.}{Figs.}
\Crefname{figure}{Figure}{Figures}
\usepackage{mathtools}
\usepackage{cuted}
\usepackage{tablefootnote}

\newcommand\revchng[1]{{#1}}

\newcommand{\acknowledgements}[1]{\begin{small}\section*{Acknowledgments}\end{small}{\noindent #1}\vspace{10pt}}
\newcommand\altaffilmark[1]{$^{#1}$}
\newcommand\altaffiltext[1]{$^{#1}$}
\newcommand\ale{\nonumber\\}

\newcommand\tnu{\nuup}
\newcommand\ad{a_{d}}
\newcommand\ai{a_{d,i}}
\newcommand\adn[1]{a_{#1}}
\newcommand\amax{a_{d,{\rm max}}}
\newcommand\amin{a_{d,{\rm min}}}
\newcommand\driftvel{{\bf w}_{s}}
\newcommand\driftveli{{\bf w}_{s,i}}
\newcommand\wsi{w_{s,i}}
\newcommand\ws{w_{s}}
\newcommand\acceli{{\bf{a}}_{i}^{\rm ext}}
\newcommand\accel{{\rm a}^{\rm ext}}
\newcommand\baccel{{\bf a}^{\rm ext}}
\newcommand\machi{\mathcal{W}_{s,i}}
\newcommand\mach{\mathcal{W}_{s}}
\newcommand\Ubar{\overline{U}}
\newcommand\hsmag{HS18}
\newcommand\hcr{H+20}

\title[]{The impact of astrophysical dust grains on the confinement of cosmic rays\vspace{-0.5cm}}

\author[Squire et al.]{
\parbox[t]{\textwidth}{ 
	Jonathan Squire\altaffilmark{1},
	Philip~F.~Hopkins\altaffilmark{2},
	Eliot~Quataert\altaffilmark{3,4},  \&\
	Philipp~Kempski\altaffilmark{3}
} 
\vspace*{6pt} \\
\altaffiltext{1}{Physics Department, University of Otago, Dunedin 9010, New Zealand} \\
\altaffiltext{2}{TAPIR, Mailcode 350-17, California Institute of Technology, Pasadena, CA 91125, USA}\\
\altaffiltext{3}{Astronomy Department and Theoretical Astrophysics Center, University of California, Berkeley, CA 94720, USA}\\
\altaffiltext{4}{Department of Astrophysical Sciences, Princeton University, Princeton, NJ 08544, USA}\vspace{-0.3cm}
}

\date{Submitted to MNRAS, ?, 2020\vspace{-0.6cm}}
\begin{document}
\maketitle

\begin{abstract}
We argue that charged dust grains could significantly impact the confinement and transport of galactic cosmic rays. 
For sub-GeV to $\sim\!10^{3}$GeV cosmic rays, small-scale parallel Alfv\'en waves, which isotropize cosmic rays through gyro-resonant interactions, are also gyro-resonant  with charged grains. If
the dust is nearly stationary, as in the bulk of the interstellar medium, Alfv\'en waves are damped by dust. This will reduce
the amplitude of Alfv\'en waves produced by the cosmic rays through the streaming instability, thus enhancing cosmic-ray transport.  
In well-ionized regions, the dust damping rate is larger by a factor of $\sim\!\!10$  than other mechanisms that damp parallel Alfv\'en waves at the scales relevant for 
$\sim\!{\rm GeV}$ cosmic rays, 
suggesting that  dust could play a key role in regulating cosmic-ray transport. In  astrophysical situations in which the dust moves 
through the gas with super-Alfv\'enic velocities, Alfv\'en waves are rendered unstable, which could directly scatter cosmic rays. 
This interaction has the potential to create a strong feedback mechanism where dust, driven through the gas by radiation pressure, then strongly 
enhances the confinement of cosmic rays, increasing their capacity to drive  outflows. This mechanism may act in the circumgalactic medium around star-forming galaxies and active galactic nuclei. 
\end{abstract}

\begin{keywords}
cosmic rays -- dust -- plasmas -- instabilities -- galaxies: evolution
\vspace{-1.0cm}
\end{keywords}

\vspace{-1.1cm}

\section{Introduction}

The mechanisms that enable and constrain the propagation and generation of cosmic rays (CRs) in galaxies remain only 
partially understood \citep{Bell2013,Amato2018}.
While it  is well accepted that, in order to explain their observed isotropy,  CRs must scatter from small-scale  
irregularities of the magnetic field, the origin and properties of these irregularities is uncertain. 
Key globally averaged quantities, such as the average CR escape time, 
can be   constrained by local observations of the CR spectrum and composition \citep[e.g.,][]{Strong1998,Ahn2010,Adriani2011,Evoli2017}.
In order to better understand these observations, it is desirable to improve our theoretical grasp of  
the physics governing particle transport across a range of energies and under
different galactic conditions. 

The problem is particularly acute for the relatively low energy $\sim\!{\rm GeV}$ protons that dominate
the CR energy density, for two  reasons. First, such CRs are 
arguably the most difficult to understand theoretically: they interact with extremely 
small $\lesssim {\rm AU}$ scales, and -- precisely because they dominate the energy density -- their transport
is likely controlled by magnetic structures that they themselves excite (making the relevant processes highly nonlinear). 
Second, and again because they dominate the energy density, $\sim\!{\rm GeV}$ CRs are thought to be important for a variety of global processes in 
galactic astrophysics, such as  launching  large-scale outflows and modifying the phase structure of the background gas \citep[e.g.,][and references therein]{Pfrommer2017,Butsky2018,Ji2020,Su2020,Hopkins2020b,Bustard2020}. This makes
understanding the details of CR transport and scattering, including the dependence on \emph{local} properties of the background
gas  (e.g., temperature, density, magnetic-field strength, and ionization fraction), particularly relevant for the $\sim\!{\rm GeV}$ CR population.
The recent study of \citet{Hopkins2020c} (hereafter \defcitealias{Hopkins2020c}{\hcr}\hcr) has highlighted this uncertainty: using cosmological 
simulations and simple scaling models, they found that none of the popular theoretical models of $\sim\!{\rm GeV}$ CR propagation could satisfactorily 
explain local grammage measurements or observations of gamma-ray emission from the halos of other galaxies   \citep{Lacki2011,Griffin2016,Lopez2018}. We are thus motivated to examine possible omissions in  current theories of CR transport.

In this work, we consider the impact of charged  dust on cosmic-ray transport. Although at first glance, dust-CR interactions
may seem esoteric and unlikely to be significant (indeed, their direct Coulomb interactions are weak; \citealp{Byleveld1993}), we find a surprisingly strong interaction that suggests dust-related effects  could 
dominate CR transport in a number  of astrophysically important regimes.
The fundamental cause of the interaction is  small-scale parallel Alfv\'en waves (AWs):
such waves are a necessary ingredient in scattering CRs and thus key to their transport; 
but they also, by very similar mechanisms,  interact strongly with charged dust. It further transpires that
 sub-AU-scale AWs, which are important to $\sim\!{\rm GeV}$ CR transport, interact with a range of grain sizes and charges that are likely 
prevalent in the interstellar medium (ISM) and circumgalactic medium (CGM). 
We suggest  two possible effects.
The most important  involves the CR ``self-confinement'' scenario, whereby CRs excite small-scale AWs through the ``streaming
instability,'' which subsequently grow in time, scattering  CRs and reducing their transport \citep{Wentzel1969,Kulsrud1969,Wentzel1974}.
In this scenario, dust acts to damp the small-scale parallel AWs excited by CRs \citep{Cramer2002}, reducing the effectiveness of 
scattering and enhancing CR transport. We find that dust-damping rates are large enough, even given the low mass fraction of dust in the ISM, 
for  the effect to dominate over all other AW-damping mechanisms in well-ionized gas. 
The second possible effect involves dust that  is moving through the gas with super-Alfv\'enic velocities due to radiation pressure 
(or some other force that affects the gas and dust differently). As studied in detail in \citet{Hopkins2018a} (hereafter \hsmag), 
AWs are unstable in this situation with a similar mechanism to the CR streaming instability; these AWs may then scatter CRs 
and enhance their confinement. Although the effect is more difficult to assess in detail than the AW damping, because 
of significant uncertainties in wave saturation physics and dust astrophysics, we speculate that it has the potential to provide 
a rather strong feedback mechanism by enhancing the coupling of CRs to  CGM gas.

We start the paper in \cref{sec: CR stuff} with an overview of the theory of CR transport and AWs, covered in sufficient detail 
to allow later discussion of the effect of dust. \Cref{sec:aw.dust.interation}, with further details in \cref{app:calc}, covers the calculation of the  
damping or growth rate of AWs in the presence of a spectrum of grain sizes. We then consider the astrophysical implications of these results
in \cref{sec:astro}, covering CR self-confinement 
in the ISM in \cref{sub: WIM}, and speculating on dust-enhanced CR confinement in the CGM in \cref{sub: CGM}. We conclude in \cref{sec:conclusion}.

\begin{table}
\begin{center}
 \begin{tabular}{||c c ||} 
 \hline
 Symbol &  Description and/or Definition  \\ [0.5ex] 
 \hline\hline
 $r_{L}$  &  CR gyroradius  \\ 
 \hline
   $ \xi=v_{\|}/v$ & CR pitch angle\tablefootnote{Most CR literature usually uses $\mu$ to denote pitch angle. We
   use $\xi$ to avoid confusion with the dust-to-gas-mass ratio, and in keeping with most plasma physics literature.} \\
 \hline
      $\tnu_{c}$  & CR scattering frequency  \\ 
  \hline
       $\kappa_{\|},\,\tilde{v}_{\rm st}$      &   CR diffusivity, ``effective'' CR streaming speed            \\
 \hline\hline
       $\rho,\,\bm{u}$      &     Gas mass density,  velocity         \\
  \hline
        $B, \,v_{A}$     &         Magnetic-field strength, Alfv\'en speed     \\            
 \hline
    $T,\,n$         &     Gas temperature, number density         \\
     \hline
    $\beta = 8\pi nk_{B} T/B^{2}$         &     Ratio of thermal to magnetic pressure         \\
     \hline\hline
    $\rho_{d},\, \bm{v}$         &     Dust mass density,  bulk velocity          \\
 \hline
        $\mu_{0}=\langle \rho_{d}\rangle/\langle \rho\rangle$     &         Total dust-to-gas-mass ratio      \\        
         \hline
        $\ad$     &         Dust-grain size    \\     
     \hline
        $\amin,\,\amax$     &         Minimum/maximum grain size in distribution    \\     
\hline
        $\xi_{\mu}$     &         Slope of dust mass distribution   \\         
\hline
        $t_{s}(\ad)$     &         Grain stopping time due to dust-gas drag   \\     
\hline
        $t_{L}(\ad)$     &         Grain Larmor time    \\    
  \hline
        $U_{d}(\ad)$     &         Grain electrostatic potential   \\    
\hline
        $\Ubar,\,\Ubar_{0}$     &        Normalized $U_{d}$  (collisional or photoelectric)  \\    
        \hline
        $\accel$     &        Grain acceleration from radiation pressure \\ 
\hline
        $w_{s}$     &        Equilibrium grain drift velocity in gas frame \\         
\hline
        $\mach=w_{s}/v_{A}-1$     &   Alfve\'enically normalized grain drift velocity      \\    
        \hline
        $\xi_{t_{L}},\,\xi_{\mach}$     &   Scaling of $t_{L}$, $\mach$ with grain size\\
  \hline\hline
        $k$     &        (Parallel) wavenumber of instability \\
         \hline
        $\omega_{A}=kv_{A}$     &        Alfv\'en frequency \\      
            \hline
        $\Gamma$     &       Instability growth or damping rate \\    
        \hline
        $k_{\amin},\,k_{\amax}$     &   Maximum, minimum unstable wavenumber \\   
 \hline       
 \end{tabular}
\end{center}
\caption{\revchng{Important symbols used throughout this article. In the linear calculations, a subscript ``$,i$'' indicates the 
quantity for grains of species/size $i$. In astrophysical estimates, we
use a numerical subscript to denote order of magnitude in cgs units, $F_{x}=F/10^{x}$; so, e.g., $T_{4}=T/(10^{4}{\rm K})$, $n_{0}=n/(1{\rm cm}^{-3})$, and $a_{-5}=\ad/(0.1\mu {\rm m})$.} }
\label{tab:}
\end{table}

 \section{Cosmic-ray scattering and self confinement}\label{sec: CR stuff}

 In this section, we outline the basic theory of Cosmic Ray (CR) scattering, briefly reviewing the relevant 
 physics. We also  point out   some interesting features related to wave polarization that 
  are specific to Alfv\'en-wave interaction with dust, exploring
 how this could affect  our later estimates of CR transport (this discussion, in \cref{subsec: CR polarization}, is unimportant to the overall narrative and may be skimmed
 without causing problems later in the text).
 
CR scattering
 is thought to occur primarily through their interaction with parallel Alfv\'en waves (AWs) with a wavelength 
 that is resonant with the distance they stream along the magnetic field in one gyro-orbit \citep{Jokipii1966}:
 $k_{{\rm res}}\xi v\approx\pm \Omega_{c}$, where $k_{\rm res}$ is the resonant wavenumber of the AW, $\xi v=v_{\|}$ is the  velocity 
  along the field line of 
a particle with pitch angle $\xi=v_{\|}/v$, and $\Omega_{c}$ is the particle's gyrofrequency. Neglecting
the $\xi$ dependence of the resonant condition and the distribution of thermal velocities, the
condition for highly relativistic protons simplifies to\begin{equation}
k_{{\rm res}}^{-1}\approx r_{L}\approx 3.3\times 10^{12}{\rm cm}\left(\frac{ R^{{\rm CR}}}{{\rm GV}}\right)\left(\frac{B}{1\mu{\rm G}}\right)^{-1}\label{eq: rL},
\end{equation}
where $B$ is the magnetic field strength, $R^{{\rm CR}}$ is the CR rigidity,   and \cref{eq: rL} applies for  $R^{{\rm CR}}\gtrsim 1{\rm GV}$.
For a given relative power $\delta \bm{B}^{2}(k_{\|})/B^{2}$ in magnetic-field fluctuations at parallel scale $k_{\|}$, 
the scattering frequency is $\tnu_{c}\sim\Omega_{c}\delta \bm{B}^{2}(k_{\|})/B^{2}$. The scattering 
brings the CRs back towards isotropy in the frame of the waves,  thus reducing the transport of CR energy.
With sufficiently efficient scattering -- i.e., with sufficient power in $r_{L}$-scale parallel waves -- the CRs behave 
like a fluid that drifts at the speed of the scatterers with respect to the gas \citep{Skilling1975,McKenzie1982,Thomas2019}. 
With less scattering, however, CRs are not efficiently isotropized, leading to significantly more energy transport. 
Thus, understanding the power in $r_{L}$-scale parallel fluctuations is crucial to understanding CR transport.
In the ISM, the $r_{L}$ scale is tiny ($\lesssim\!{\rm AU}$) for the $\sim\!{\rm GeV}$ CRs that dominate the energy density, and 
 although interstellar turbulence is expected to put significant power into \emph{perpendicular}
Alfv\'enic fluctuations on such scales, perpendicular fluctuations scatter CRs very inefficiently  
and likely cannot account for the observed isotropy of CRs (\citealp{Chandran2000,Yan2002}; \hcr). This suggests that 
other sources of small-scale parallel AWs are needed. Note that CRs can, in principle, scatter from non-Alfv\'enic 
fluctuations also, but such fluctuations are strongly damped at $r_{L}$ scales
and there remains debate as to whether they can play a significant role (\citealp{Yan2004a,Yan2008}; \hcr).

Given this apparent inability of turbulence to provide sufficient  CR scattering, the most commonly 
accepted source of $r_{L}$-scale fluctuations is the CRs themselves. 
 In the standard ``self-confinement'' picture of CR transport \citep{Wentzel1969,Kulsrud1969,Skilling1971}, 
streaming CRs
 generate parallel  AWs due to the ``streaming instability'' whenever their 
bulk drift velocity exceeds the local Alfv\'en speed of the background plasma. The waves, with wavelengths 
that automatically match the CR gyroradii, then efficiently scatter particles, bringing the CR bulk drift speed back towards the Alfv\'en speed. 
If undamped, it is expected that such AWs would grow to  large amplitudes, 
causing efficient scattering of CRs and bringing their drift back to nearly exactly the Alfv\'en speed \citep{Skilling1975,Skilling1975b}. In contrast, 
if there exists an efficient mechanism to damp parallel AWs, the CR-AW system is expected to come
to a local equilibrium where the growth rate of AWs due to the CR streaming is balanced 
by damping, implying that the CR energy density can be transported at speeds significantly 
faster than the Alfv\'en speed. Thus, in the self-confinement paradigm, an understanding of parallel AW damping is crucial 
to understanding CR transport: more efficient damping leads to faster CR transport.
Current understanding suggests that ion-neutral viscosity \citep{Kulsrud1969},   damping from the 
interaction of parallel Alfv\'en waves with turbulence  \citep{Farmer2004,Zweibel2017}, 
and  nonlinear Landau damping \citep{Lee1973,Cesarsky1981,Voelk1982}, 
are the most important physical processes. In predominantly neutral gas (e.g., cold clouds, the warm neutral 
medium), ion-neutral damping is highly dominant and leads to rapid transport of CRs. However, 
as the ionization fraction increases, ion-neutral damping rates quickly drop to zero. Through  galaxy-scale simulations
and simple analytic arguments, \hcr\ argued that 
it is transport through well-ionized gas that dominates the global confinement properties, while 
fast damping rates (fast transport) in predominantly neutral regions make little difference \citep[see also][]{Farber2018}. Further, \hcr\ found that  current self-confinement transport models   predict too much CR confinement in ionized gas  to match observations.

We are thus left with two possible roles for dust in regulating CR transport. In the first -- ``external confinement'' -- drifting dust may directly excite small-scale parallel AWs due to instability. Such 
waves, if they can reach large amplitudes, will scatter CRs directly, thus increasing their confinement. In the second -- ``self-confinement'' -- near-stationary dust will act to damp small-scale AWs excited by the CRs themselves, thus decreasing
their confinement.

\subsection{Polarization}\label{subsec: CR polarization}

An interesting feature of dust-AW interaction, which may have consequences for both 
external confinement and self confinement of CRs, is the dependence on wave polarization. 
Specifically, CRs with a given sign of $v_{\|}$ interact only with one AW polarization:  \revchng{right-handed waves propagating parallel (or left-handed 
waves propagating anti-parallel)
to the  magnetic field scatter only  $v_{\|}>0$ positively charged CRs, while the opposite is true for  $v_{\|}<0$ CRs}. The CR streaming 
instability excites nearly linearly polarized AWs because the populations of $v_{\|}<0$ and $v_{\|}>0$ particles
are nearly the same \citep{Kulsrud1969,Bai2019}; however, only waves that propagate in the CR-drift direction are
unstable. For  similar reasons, AW damping or instability due to dust is also dependent on wave polarization, 
in contrast to the other wave damping mechanisms (e.g., turbulence damping) mentioned above. \revchng{In particular
(see \cref{sec:aw.dust.interation} below), for waves propagating along the magnetic-field direction, negatively (positively) charged stationary  dust  
damps only right-handed (left-handed) waves; vice-versa for waves propagating in the anti-field direction. Similarly, dust drifting super-Alfv\'enically in the 
magnetic-field direction excites 
left-handed waves if negatively charged, or  right-hand waves if positively charged, and in both cases only 
waves propagating in the same direction as the dust are unstable. }
In this section we consider how damping or growth of just one polarization of AW 
would affect CR self confinement and scattering. We wish to understand how to appropriately compare 
AW damping or growth rates from dust with those due to other mechanisms.

First consider the self-confinement scenario, with dust providing the dominant damping mechanism
for nearly linearly polarized AWs excited by the CR streaming instability. As noted above, 
dust (if negatively charged) will damp only the right-handed (\revchng{forward propagating}) waves, thus leaving
left-handed waves to grow freely to larger amplitudes. There will thus be efficient scattering of 
$v_{\|}<0$ particles, which will tend to make the CR distribution function independent of $\xi$
for $v_{\|}<0$; but, the $v_{\|}>0$ part of the CR distribution function, which only interacts with the strongly damped waves, remains nearly unaffected. 
However, because the overall  drift velocity of the CRs arises from the difference in 
the total population of $v_{\|}<0$ and $v_{\|}>0$ particles, flattening of the $v_{\|}<0$ part
of the distribution in $\xi$ makes only a minor difference to the CR drift and energy transport, or to the streaming 
instability growth rate
(the effect is nicely illustrated in \citealt{Bai2019} figure 9). Further, it seems likely 
that isotropization of particles across  $\xi=0$  (often called the ``$90^{\circ}$ barrier'')
requires variation in $|\bm{B}|$, so would not be efficient in the presence of only one wave polarization \citep{Felice2001,Bai2019,Holcomb2019}.
Overall, this suggests that the effective transport will be approximately  determined by the level of the \emph{lowest} amplitude waves
(right-handed in the discussion above), with little dependence on a possibly large-amplitude population 
of the other polarization. Thus, damping just one wave polarization through interaction
with dust should have a similar effect on CR self confinement to damping both polarizations, implying it is reasonable
to compare dust AW damping rates directly to those from other sources that do not distinguish between polarizations (e.g., turbulence damping).
Of course, there exist a variety of complexities in the above argument that make it highly uncertain; however, 
it seems likely that such questions may only be answered definitively with detailed simulations of the streaming instability, which are only  
recently becoming possible \citep{Bai2019,Holcomb2019,Dubois2019,Haggerty2019,Weidl2019}.

From the above argument, we also see that Alfv\'en waves that are  excited directly by drifting dust (``external confinement'')
may be rather inefficient at confining CRs. Before the instability saturates, and neglecting 
other sources of AWs including those excited by CRs, only one polarization of AW will be present. Such a spectrum of AWs 
 will not significantly perturb the magnetic field strength  and will only scatter CRs with one sign of $\xi$. 
Such a process can only limit the energy transport of CRs  by a maximum factor of $\sim\!2$.
 This effect was seen and diagnosed in detail in the CR streaming-instability simulations of \citet{Holcomb2019}, where they noted that
 CRs with an initial drift velocity that is too large (approaching $c$) are inefficiently self confined, because the wave scattering only isotropizes particles with one sign of $\xi$. 
 However, the nonlinear saturation of the dust instability presents significant uncertainties; if the saturation 
 is quasi-turbulent in nature, as suggested by the simulations of \citet{Seligman2019,Hopkins2020} (albeit in a  different regimes), it seems 
 implausible that only one wave polarization would be present in the saturated state. In that
 case, fluctuations induced by the dust could more efficiently isotropize CRs.
There are clearly additional uncertainties, such as how different  spatial regions behave when there is a large-scale gradient of CRs,
 and detailed simulations of both the saturation of the dust instability and the CR scattering are needed for 
 better understanding of the system. Overall, however, it seems 
reasonable to surmise that CR confinement from small-scale AWs excited by dust  could be  less efficient 
than suggested by  estimates of $\delta \bm{B}^{2}(k_{\|})/B^{2}$ alone.

\section{The interaction of highly charged dust with Alfv\'en waves}\label{sec:aw.dust.interation}

The goal of this section is to compute the damping or growth rate of parallel shear-Alfv\'en waves 
in the presence of a wide spectrum of grain sizes. Our method involves first computing the dispersion relation
of Alfv\'en waves in the presence of  a discrete set of grain sizes
in the relevant low dust-to-gas-mass-ratio limit (see also \citealt{Tripathi1996,Cramer2002}). A surprisingly simple expression
for the continuum system is obtained by taking the limit as the total number of grain species approaches 
infinity, while keeping the total mass density of dust constant. Importantly, this expression 
scales with the total dust density, is effectively independent of the dust drag time, and converges 
 rapidly to the continuum limit, implying that large damping and growth rates occur even with a 
continuum of grain sizes (unlike, e.g., the low dust-to-gas-ratio streaming instability; \citealp{Krapp2019}). 
\revchng{A simple extension of the calculation shows that an (isothermal)  dust pressure response does not
change the damping or growth rate, suggesting that a dust velocity dispersion -- driven, e.g., by gas turbulence \citep{Yan2004} -- 
will not strongly affect the shear-Alfv\'enic modes of interest (although a true kinetic treatment is needed to formally probe this physics; see \cref{app:subsub vel dispersion}).}
Mathematical details of the calculation are given in \cref{app:calc}.

\subsection{Dust model and definitions}\label{sub: aw dust setup}

We model dust as a charged, pressureless fluid that interacts with the gas through drag
and Lorentz forces. The gas is modelled with the ideal MHD equations, which 
is also a reasonable model for a collisionless plasma for the scales and modes 
being considered here \citep{Schekochihin2009}. Our equations do, however, assume that the total charge
contained in the dust species is small compared to that of the background gas \citep{Shukla2002}. For a set of $N_{d}$ dust species, labeled $i$, the equations are 
\begin{align}
\frac{\partial\rho}{\partial t} & + \nabla\cdot (\bm{u} \rho) = 0, \label{eq:MHD.dust.rho} \\
\frac{\partial\rho_{d,i}}{\partial t} &+ \nabla\cdot (\bm{v}_{i} \,\rho_{d,i}) = 0,\label{eq:MHD.dust.rhod}  \\
\frac{\partial\bm{u}}{\partial t} + \bm{u}\cdot\nabla\bm{u}  = &\bm{g} -\frac{1}{\rho}\nabla \left( p + \frac{B^{2}}{8\pi}\right)- \frac{\bm{B}\cdot\nabla\bm{B}}{4\pi \rho}, \ale
& + \sum_{i=1}^{N_{d}} \frac{\rho_{d,i}}{\rho}\left( \frac{\bm{v}_{i}-\bm{u}}{t_{s,i}} - \frac{\bm{v}_{i}-\bm{u}}{t_{L,i}} \times \hat{\bm{b}} \right)\label{eq:MHD.dust.u}  \\
\frac{\partial\bm{v}_{i}}{\partial t} + \bm{v}_{i}\cdot\nabla\bm{v}_{i}  = &\bm{g} +\acceli + \left( \frac{\bm{u}-\bm{v}_{i}}{t_{s,i}} - \frac{\bm{u}-\bm{v}_{i}}{t_{L,i}} \times \hat{\bm{b}} \right),\label{eq:MHD.dust.v} \\
\frac{\partial\bm{B}}{\partial t} + \bm{u}\cdot\nabla\bm{B} = &\bm{B}\cdot\nabla\bm{u} - \bm{B}\nabla\cdot\bm{u} .
\label{eq:MHD.dust.b} 
\end{align}
Here $\rho$, $\bm{u}$, $p$, and $\bm{B}$ are respectively the gas density, velocity, pressure, and the magnetic field, while $\hat{\bm{b}}$ is the magnetic field unit vector ($\bm{B}=B\hat{\bm{b}}$) and $\bm{g}$ is an external gravitational force on both gas and dust. We also define the Alfv\'en speed $v_{A}=B/\sqrt{4\pi \rho}$.
The continuum mass density and bulk velocity of dust species $i$ are  $\rho_{d,i}$ and $\bm{v}_{i}$ respectively, and $\acceli$
is an external force that can act differently on each dust species (e.g., due to radiation pressure). The  physics of a particular dust species 
is determined by   its microscopic parameters: the
 grain radius $\ai$, mass $m_{d,i}$,  solid density $\bar{\rho}_{d,i}=3m_{d,i}/(4\pi \ai^{3})$,   charge $q_{d,i}=Z_{d,i}e$, and electrostatic potential $U_{d,i}\approx q_{d,i} /\ai$. 
 These parameters determine the stopping time $t_{s,i}$, which is also  a function of gas parameters (see \cref{subsub: drag}), and the 
Larmor time 
\begin{equation}
t_{L,i}=\frac{ m_{d,i} c}{q_{d,i} B},\label{eq: tL definition}
\end{equation}
\revchng{which can be either positive or negative depending on the sign of $q_{d,i}$.}
We use $\langle \cdot\rangle$ to denote an equilibrium (background) quantity, and define the dust-to-gas mass ratio $\mu_{i}=\langle \rho_{d,i}/\rho\rangle$,
and the total dust-to-gas-mass ratio $\mu_{0}=\sum_{i}\mu_{i}$. 

As shown in \hsmag, in the presence of an external force on the dust ($\acceli\neq0$), a formal quasi-equilibrium is set
up in which both dust and gas accelerate at the same rate, but with some velocity offset, denoted $\driftvel$. Moving
into the frame in which the gas is stationary allows one to study instabilities 
about the equilibrium $\langle \bm{v}_{i}\rangle=\driftveli$, $\langle \bm{u}\rangle=0$, with the free-energy source for the instabilities
arising from the net drift of each species of dust. 
For arbitrary $t_{s,i}/t_{L,i}$, and when the  angle between $\acceli$ and the background magnetic field $\bm{B}_{0}$ is arbitrary, 
the expression for $\driftveli$ is rather complex, arising from the balance between magnetic and drag forces on grains. 
\revchng{However, as shown by \hsmag\ (see their equation 2), in the limit   $|t_{L,i}|\ll t_{s,i}$, $\driftveli$ tends to align more and more closely 
with $\bm{B}_{0}$, albeit with a magnitude that is reduced by the
projection factor $(\acceli\cdot \bm{B}_{0})/(|\acceli|| \bm{B}_{0}|$).} Intuitively, this corresponds
to the fact that well magnetized particles are only free to move along the magnetic field direction.
Because most regimes of interest for the study of CR propagation satisfy $|t_{L,i}|\ll t_{s,i}$ (see \hsmag\ figure 6), 
we thus assume that $\driftveli$ lies parallel to $\bm{B}_{0}$, simplifying the analysis enormously.  \revchng{The general case where
$\acceli$ and $\bm{B}_{0}$ are not parallel is thus effectively contained within our analysis by including some order-unity 
 projection factor $\Phi\equiv (\acceli\cdot \bm{B}_{0})/(|\acceli|| \bm{B}_{0}|)$, but without having to account for a complex, three-dimensional equilibrium where dust drifts 
at an arbitrary angle to the magnetic field. So long
as $|t_{L,i}|\ll t_{s,i}$, this is a good approximation on timescales longer than $\sim\!t_{s,i}$ (the time taken to for the system to reach equilibrium). }

\subsection{Damping and growth of parallel Alfv\'enic modes}\label{sub: growth rate calc}

Our analysis proceeds in the standard way by linearizing  \crefrange{eq:MHD.dust.rho}{eq:MHD.dust.b} about the quasi-equilibrium 
described above: $\langle \bm{u}\rangle=0$, $\langle \rho\rangle=\rho_{0}$,  $\langle \bm{v}_{i}\rangle=\wsi\hat{\bm{z}}$, $\langle \rho_{d,i}\rangle=\mu_{i}\rho$, $\langle \bm{B}\rangle=\bm{B}_{0}=B_{0}\hat{\bm{z}} $, where $w_{s,i}$ can be zero
(stationary dust). We then insert the Fourier ansatz for the evolution of 
each linearized quantity -- $\delta f(\bm{x},t) \equiv f(\bm{x},t)-\langle f\rangle= \delta f\exp(i\bm{k}\cdot \bm{x}-i\omega t)$ for $f= \bm{u},\,\rho,$ etc. -- to convert  \crefrange{eq:MHD.dust.rho}{eq:MHD.dust.b} into an
eigenvalue equation for the frequency $\omega$.  $\Im(\omega)<0$ ($\Im(\omega)>0$) implies that a particular mode is damped
(growing).
A further significant simplification comes from specializing to purely parallel modes $\bm{k}=k\hat{\bm{z}}$ ($k_{x}=k_{y}=0$).
The justification for this simplification is  that CRs  interact strongly only with purely parallel modes \citep{Kulsrud1969,Chandran2000},
so even if an oblique mode were to grow (or be damped) more rapidly, there will be little effect on CR scattering. In addition, 
parallel modes are usually the fastest growing in the $|t_{L}|\ll t_{s}$ regime (see \hsmag).

With these simplifications, it transpires that one can obtain accurate, simple forms for the dispersion relation 
$\omega(k)$ for modes with frequencies near the Alfv\'en frequency $\omega_{A} = k v_{A}$. Specifically, one inserts the ansatz $\omega = kv_{A}+\mu(\omega^{(1)}/\mu)$ (such 
that $\omega^{(1)}$ is the perturbed frequency), and expands the characteristic equation in the small parameter $\mu\ll1$ with $\omega^{(1)}/\mu$ finite, taking  $\mu_{i}=\bar{\mu}_{i}\mu$ (i.e., all individual dust densities $\mu_{i}$ scale with $\mu$).
Solving the  polynomial equation 
for $\omega^{(1)}$ that appears at lowest order in $\mu$ and taking its imaginary part leads to the  result
\begin{align}
\Im(\omega^{(1)})=&\omega _A \sum_{i=1}^{N_{d}}\mu_{i}\tilde{\omega}^{(1)}_{i}\ale 
=\omega_{A}\sum_{i=1}^{N_{d}}&\mu_{i}\machi^2  \bar{t}_{s,i} \frac{\machi^2 \bar{t}_{s,i}^2\left(2\bar{\lambda }_{i}^2 \machi+\bar{\lambda }_{i}^{2}-1\right)
   -1}{(\bar{\lambda }_{i}^2-1)^2 \machi^4 \bar{t}_{s,i}^4+2 (\bar{\lambda }_{i}^2+1)\label{eq:discrete.growth.rate}
   \machi^2 \bar{t}_{s,i}^2+1}.
\end{align}
Here, $\machi\equiv \wsi/v_{A}-1$ is the relative Alfv\'en Mach number of the streaming dust and $\bar{t}_{s,i}=\omega_{A}t_{s,i}$ is
the normalized stopping time. 
 $\bar{\lambda }_{i}=k^{-1}/\lambda_{{\rm res},i}$ is the mode's inverse wavenumber normalized by the resonance wavelength $\lambda_{{\rm res},i}\equiv v_{A}t_{L,i}\machi$, which is the  inverse wavenumber of the Alfv\'en wave mode that matches the 
 streaming gyro-orbit frequency of the dust; i.e., the wavelength for which $\omega_{A}=k w_{s,i} - t_{L,i}^{-1}$. 
 As expected, the contribution of each grain species ($\tilde{\omega}^{(1)}_{i}$ in  \cref{eq:discrete.growth.rate}) is negative (damping)
 if $\machi<0$ and positive (unstable) if $\mach>0$.

To take the continuum limit  of  \cref{eq:discrete.growth.rate}, we note that $\tilde{\omega}^{(1)}_{i}$ becomes increasingly sharply peaked around the resonance ($\bar{\lambda}_{i}=1$) at increasing $\bar{t}_{s,i}\gg1$.
This implies that each species contributes to the total  damping or growth rate only around the resonant wavenumber  $\bar{\lambda}_{i}=1$. As shown in \cref{app:calc}, this  allows for the straightforward derivation of a simple expression for continuum damping or growth rate, 
\begin{equation}
\Gamma_{\rm dust}(k) =\sgn(\mach)\left.\omega_{A}\mu_{0}\frac{\pi}{2}\frac{d\bar{\mu}(\ad)}{d\ln \ad} \frac{ \mach(\ad)^{2}}{|\xi_{t_{L}}+\xi_{\mach}|}\right|_{\,k^{-1} = v_{A}t_{L}(\ad)\mach(\ad)}.\label{eq:growth.rate} 
\end{equation}
Here, we have changed from labelling each dust species by its discrete index $i$, to making $\mach$ and $t_{L}$ functions of physical 
grain size $\ad$, defining \begin{equation}
\xi_{t_{L}}\equiv\frac{d\ln |t_{L}|}{d\ln \ad},\qquad \xi_{\mach}\equiv\frac{d\ln |\mach|}{d\ln \ad}.\label{eq: xi tL xi w defs}
\end{equation}
The fractional mass across a given range of sizes is parameterized by $d\bar{\mu}/d\ln \ad$, which satisfies $\int d\ln{\ad}\,(d\bar{\mu}/d\ln \ad)=1$ across the full range of grain sizes present (i.e., $d\bar{\mu}/d\ln \ad$ is the fractional contribution to dust density 
from grains of size $\ad$).\footnote{For the standard MRN size distribution, $dn_{d}\propto a^{-3.5} d\ad$ \citep{Mathis1977}, $d\bar{\mu}/d\ln \ad \propto \ad^{0.5}$.   } Evaluating  \cref{eq:growth.rate} as a function of $k$  involves first inverting the
resonance condition, $ v_{A}t_{L}(\ad)\mach(\ad)=k^{-1}$,  to find $\ad(k)$, then inserting this into the main expression \eqref{eq:growth.rate}. 
More detail about the derivation of  \crefrange{eq:discrete.growth.rate}{eq:growth.rate} and how they relate to the
 Resonant 
Drag Instability theory of \cite{Squire2018} and \hsmag\ is given in  \cref{app:calc}. We also confirm 
 the validity of  \cref{eq:growth.rate} in  \cref{fig:app:growth.rate}, by comparing to numerical solutions of the 
 full dispersion relation for a discrete set of dust grains.
 
 \subsubsection{Polarization}\label{subsub: polarization}
 
Our derivation of  \cref{eq:discrete.growth.rate} from the frequency alone has hidden the 
relevance of the mode polarization inside the sign of $t_{L}$, which controls the sign of $k$ (or $\bar{\lambda}_{i}$) and thus the
mode propagation direction and polarization. Physically, only modes that resonate with 
the dust can interact with it, which, for negatively charged dust \revchng{and waves propagating in the magnetic-field direction}, implies that only right-hand polarized 
waves are damped for $\mach<0$, while only left-hand polarized waves are unstable for $\mach>0$. 
The opposite is true for positively charged dust \revchng{and/or for wave propagation in the anti-magnetic-field direction.}
The handedness of the damped/unstable waves (determined by the dust charge) does not feature 
prominently in the discussion below, because the streaming speed of CRs ($\sim\! v_{A}$) 
is generally much smaller than the speed  of individual particles ($\sim\!c$), implying there 
are nearly equal numbers of forward and backward propagating particles. 
However, the fact that only one polarization is damped or unstable does suggest 
some interesting implications on CR transport, as discussed in \cref{subsec: CR polarization}.

\section{Astrophysical consequences}\label{sec:astro}

In this section, we consider how the dust-induced damping or growth of parallel Alfv\'en waves (AWs)  could 
impact CR confinement in galaxies. We suggest two possible effects, which have opposite consequences for
CR propagation. The first -- discussed in \cref{sub: WIM} -- involves the additional AW damping caused by dust 
reducing the efficiency of CR self-confinement,  thus enhancing the CR transport. The second -- discussed in \cref{sub: CGM} -- considers how dust with super-Alfv\'enic drift speeds in the circumgalactic medium (CGM) could excite parallel AWs that directly 
scatter CRs, thus enhancing the  CR confinement (reducing the transport). We start with a brief review of the dust properties
that will be necessary for our discussion, focusing on expressions and/or physical processes that are relevant to $\ad\lesssim 1\mu\rm{m}$ grains in well-ionized gas with a temperature $T\gtrsim 10^{4}\rm{K}$. This 
focus is motivated by the fact that in colder, predominantly neutral gas, 
CRs are thought to diffuse rapidly anyway due to strong 
ion-neutral damping, which will generally dominate the effects we discuss here for reasonable dust-to-gas-mass ratios.

In addition 
to quantities defined above, throughout this section it will be convenient to use subscripts to denote orders of magnitude in cgs units with $F_{x}=F/10^{x}$ for 
some quantity $F$; thus  $T_{4}$ is 
the gas temperature in units of $10^{4}{\rm K}$, $n_{0}$ is the gas number density in units of ${\rm cm}^{-3}$, $B_{-6}$ is the magnetic-field strength in units of $\mu{\rm G}$,  $\adn{-5}$ is the grain radius $\ad$ in units of $0.1 \mu\rm{m}$, and $\bar{\rho}_{d;0}$ is the grain solid density in units of $\rm{g}\,\rm{cm}^{-3}$. We also define 
the plasma ``beta,'' which is the ratio of thermal pressure to magnetic pressure $\beta=8\pi P/B^{2}$.


\subsection{Physical properties of astrophysical dust}\label{sub: dust stuff}

The key dust properties of interest for computing the damping/growth rate of Alfv\'en waves from  
 \cref{eq:growth.rate}
are: (i) the dust drag law $t_{s}(\ad)$, which, in addition to the external force on the grains, determines $\mach(\ad)$; (ii) the dust charge, which 
determines $t_{L}(\ad)$ from  \cref{eq: tL definition}; and (iii) the mass distribution of grain sizes $d\bar{\mu}/d\ln a$. Let
us discuss each of these in turn, summarising relevant information from previous literature.

\subsubsection{Drag law}\label{subsub: drag}
The two relevant expressions for grains in the conditions of interest here are Epstein drag, which is collisional drag when the particle
size is smaller than the gas mean free path, and Coulomb drag, which arises when charged grains interact with the
background plasma \citep{Draine1979a}.
The stopping time for grain species $i$ in the Epstein  regime is approximately 
\begin{align}
t_{s,i}^{\rm Ep}&= \sqrt{\frac{\pi }{8}}\frac{\bar{\rho}_{d,i} \ai}{\rho\, v_{\rm th}} \left(1+ \frac{9\pi }{128}\frac{|\bm{v}_{i}-\bm{u}|^{2}}{v_{\rm th}^{2}}\right)^{-1/2}\nonumber\\
&\approx 6.8\times 10^{12}{\rm s} \frac{\adn{-5}\bar{\rho}_{d;0}}{n_{0}T_{4}^{1/2}},
\label{eq: epstein drag}
\end{align}
where $v_{\rm th}$ is the gas thermal velocity, 
while in the Coulomb regime it follows
\begin{align}
t_{s,i}^{\rm Coul}&\approx \sqrt{\frac{\pi}{2}}\frac{\bar{\rho}_{d,i}\ai }{f_{\rm ion} \rho\, v_{\rm th} \ln \Lambda_{d}  } \left(\frac{k_{B} T}{z_{i}e U_{d}}\right)^{2} \left(1+ \sqrt{\frac{2}{9\pi}} \frac{|\bm{v}_{i}-\bm{u}|^{3}}{v_{\rm th}^{3}}\right),\nonumber\\
&\approx 1.1\times 10^{11}{\rm s} \frac{\adn{-5}\bar{\rho}_{d;0}}{f_{{\rm ion}}n_{0}T_{4}^{1/2} \Ubar^{2}},
\label{eq: coulomb drag}
\end{align}
where $f_{\rm ion }$ is the ionization fraction, $\ln \Lambda_{d}$ is the Coulomb logarithm for the dust ($\ln \Lambda_{d} \approx20$ for the conditions and grains of interest; see \hsmag; \citealt{Draine1979a}), $z_{i}$ is the mean gas ion charge, and 
$-2.5 \Ubar=U_{d}/(k_{B}T/e)\approx e^{2} Z_{d}/(\ad k_{B}T)$ is the normalized grain potential (see \cref{subsub:grain charge} below).
In the second lines of  \crefrange{eq: epstein drag}{eq: coulomb drag} we have assumed subsonic grain drift $ |\bm{v}_{i}-\bm{u}|\ll v_{{\rm th}}$,  taken the 
mean molecular weight to be 0.6 (a typical choice for warm ionized gas), and taken $z_{i}\approx 1$.
The correct drag law is approximately whichever 
of  \cref{eq: epstein drag,eq: coulomb drag} is smaller. For maximally collisionally  
charged ($\Ubar\approx1$) grains moving subsonically in well-ionized gas, this is Coulomb drag, 
otherwise Epstein drag dominates.

\subsubsection{Grain charge}\label{subsub:grain charge}

Grain charging processes are very complicated  and uncertain, being strongly influenced  
by a range of environmental effects and grain microphysics. Further, in
many systems there is expected to be a wide distribution of different
grain charges for a given grain size. Here we use several different simple analytic fits to 
approximate the average charge expected in different regimes, as discussed in, e.g.,  \citet{Draine1979a,Draine1987,Weingartner2001b,Weingartner2004,Tielens2005}.

The simplest relevant process is collisional charging, which satisfies approximately
\begin{equation}
U_{d} \approx \frac{e Z_{d}}{\ad}\approx -\frac{e}{\ad}\frac{1}{1+0.037\sqrt{\frac{e^{2}}{ak_{B}T}}} - 2.5 \frac{k_{B}T}{e}\approx - 2.5 \frac{k_{B}T}{e},\label{eq:grain charge collisional}
\end{equation}
where the latter approximation is valid for the conditions of interest ($T\gtrsim 10^{4}{\rm K}$).
This motivates defining $-2.5\Ubar=U_{d}/(k_{B}T/e)$, with $\Ubar\approx1$ for collisionally charged grains. For reference, 
the grain potential is $U_{d}\approx -2.15 T_{4}\Ubar\,{\rm V}$.
When $Z_{d}$ becomes too negative, the total charge becomes limited 
by electron field emission \revchng{to $Z_{d}\approx -7000 \adn{-5}^{2}$, which} translates to 
\begin{equation}
\Ubar\approx \Ubar_{0}\min  \left\{1, 47 \frac{\adn{-5}}{T_{4}}\right\},\label{eq: grain charge no photoelectric}
\end{equation}
where we retain a dimensionless $\Ubar_{0}\approx1 $ factor, so as to understand the charge dependence of analytic expressions that we later derive. \revchng{At 
higher gas temperatures  ($T\gtrsim 10^{5}{\rm K}$ for silicate grains, $T\gtrsim 10^{5.5}{\rm K}$ for graphite), secondary electron emission can 
cause grains to become positively charged. This process can be important, especially for smaller grains where it can lead to larger
$|U_{d}|$ than the standard collisional expression \eqref{eq: grain charge no photoelectric}; but,
we do not consider it in detail
because of the complex, material-dependent expressions (see figures 1 and 2 of   \citealt{Draine1979a}).}

In the presence of a  radiation field, photoelectric charging can dominate, which 
also causes grains to gain net positive charge. The process depends primarily on the gas/radiation through the parameter, $\psi=G_{0} T^{1/2}/n$, where 
$G_{0}=u^{{\rm uv}}_{{\rm rad}}/(5.3\times 10^{-14} {\rm erg}\,{\rm cm}^{-3})$ and $u^{{\rm uv}}_{{\rm rad}}$ is the energy density in the radiation field between $\sim \!6{\rm eV}$ and $13.6{\rm eV}$. 
The charge is approximately $Z_{d}\approx 36 (\psi/1000) \adn{-5}$ up to a maximum potential $U\sim7{\rm V}$ or $Z_{d}\approx 500 \adn{-5}$  (see \citealt{Weingartner2001b,Tielens2005,Draine2010} and figure 4 of \citealt{Weingartner2004}).
This gives 
\begin{equation}
\Ubar\approx -\Ubar_{0}T_{4}^{-1}\min\left\{ 0.24 \left(\frac{\psi}{1000}\right),\,3.3 \right\}.\label{eq: grain charge photoelectric}
\end{equation}
Overall, we see that in most regimes, $\Ubar$ may be assumed independent of $\ad$ in 
order to compute the $\mach$ and $t_{L}$ derivatives required to evaluate damping/growth of Alfv\'en waves (\cref{eq:growth.rate}), while for small grains in hot gas without a strong radiation field, $\Ubar\propto \ad$. 
For simplicity, we will neglect the quantization of grain charge in analytic estimates, although this does become significant for the smallest grains (see, e.g., figure 25.3 of \citealt{Draine2010}).

\subsubsection{Mass distribution}\label{subsub:mass distribution}

Grain mass distributions probably 
vary significantly between regions, depending on the complex interplay between 
grain growth, grain shattering through collisions, and grain sputtering from the gas \citep{Peters2017}. The standard MRN distribution of \citep{Mathis1977}
postulates that $d\bar{\mu}/d\ln \ad\propto \ad^{0.5}$ with $\amin\approx 5 {\rm nm}$ ($a_{-5}=0.05$) and 
$\amax\approx 0.25 \mu{\rm m}$ ($a_{-5}=2.5$), and a total dust to gas mass ratio of around $1\%$ ($\mu_{0}=0.01$).
The MRN distribution likely has significant inaccuracies even in the  ISM,
missing a sizeable population of small grains with $\ad<5{\rm nm}$  \citep[e.g.,][]{Weingartner2001,Zubko2004,Draine2009}.
Far less is known about the CGM or hotter regions. Grain destruction due to ion-field emission is expected to be significant only  for very small grains in very hot gas, and not strongly affect our results here (e.g., in $T\sim 10^{6}{\rm K}$ gas,  \citealt{Draine1979a} suggest the ion-field-emission-limited grain size is $a\lesssim 1{\rm nm}$). Thermal sputtering rates become   larger than grain growth due to accretion of metals from the gas 
for $T\gtrsim 10^{5}{\rm K}$, and significant compared to gas dynamical times for $T\gtrsim 10^{6}{\rm K}$ \citep{Draine1979,Draine1979a}, suggesting that the dust-to-gas mass ratio may be  lower in hotter gas.
However, dust is clearly observed in the  CGM of galaxies   \citep{Menard2010,Peek2015} and thus appears to 
survive in gas up  to $T\sim 10^{7}{\rm K}$, although this may also be related to the presence of multiphase gas (i.e., 
dust existing primarily in clumps of colder gas; \citealp{Tumlinson2017}). While its size distribution remains very uncertain,  the presence 
of smaller  grains (down to at least  $\ad\sim 0.01\mu{\rm m}$) is indicated by observations \citep{Hirashita2020}.

Given 
these very significant uncertainties, the 
use of a complex dust-size distribution model would be of dubious value,\footnote{Further,  the
 mass distribution of grains is important
for wave damping/growth only insofar as it determines the
mass density of grains with a given $t_{L}$, $\mach$, and $t_{s}$. This implies that  the mass 
distribution gets mixed with the charge distribution, which has its own significant uncertainties.}  and for simplicity  we will use a power-law grain mass distribution $d\bar{\mu}/d\ln \ad\propto \ad^{\xi_{\mu}}$ between $\amin\approx 1 {\rm nm}$ and $\amax\approx  0.25 \mu{\rm m}$ in all estimates, with $\xi_{\mu}=0.5$ (MRN) as the fiducial choice.
Since wave growth/damping rates around a particular wavelength scale linearly with the density of dust that is resonant with that wavelength, 
one can simply adjust the total dust-to-gas mass ratio   $\mu_{0}$ to account for uncertainty in the mass distribution.


\subsection{Enhanced cosmic-ray transport in the Warm Ionized ISM}\label{sub: WIM}

\begin{figure*}
\begin{center}
\includegraphics[height=0.755\columnwidth]{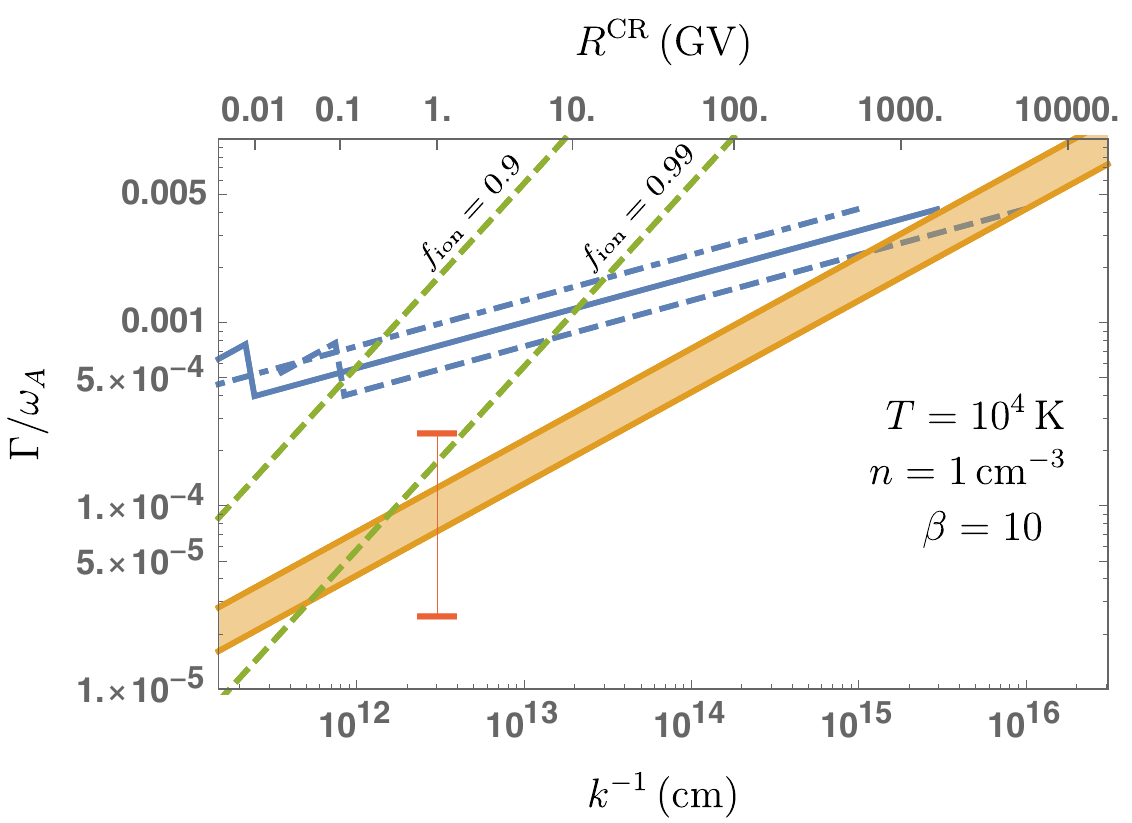}\includegraphics[height=0.74\columnwidth]{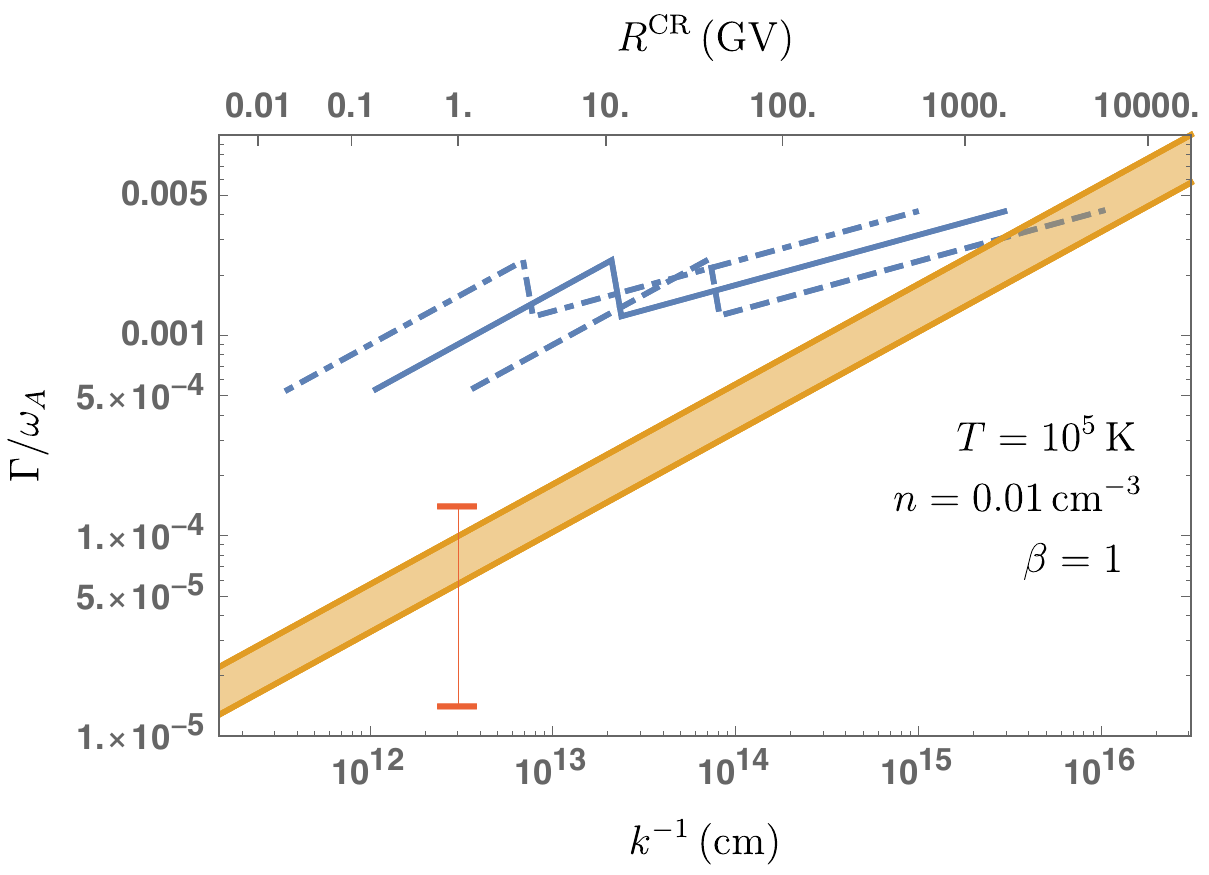}
\caption{Comparison of parallel AW damping due to dust (blue lines;  \cref{eq: WIM damping}), turbulent 
damping (yellow region;  \cref{eq: turbulent damping}), nonlinear Laudau damping (red bar;  \cref{eq: NLL damping}), and ion-neutral damping (green line;  \cref{eq: in damping}) as a function of scale. The top axis shows the 
rigidity of CRs that are resonant with AWs at $k^{-1}$ for the conditions of interest. The left-hand panel illustrates cooler, denser
conditions, with $T=10^{4}{\rm K}$, $n=1{\rm cm}^{-3}$, and $\beta=10$, while the right-hand panel illustrates hotter, more diffuse 
conditions $T=10^{5}{\rm K}$, $n=0.01{\rm cm}^{-3}$, and $\beta=1$. For the dust, the total dust to gas mass ratio is $\mu_{0}=0.01$, and grains range between $\amin=0.01\mu{\rm m}$, $\amin=0.25\mu{\rm m}$ with a mass distribution $d\bar{\mu}/d\ln a\propto a^{0.5}$. We consider collisional charging  (\cref{eq: grain charge no photoelectric}), and the different lines show the effect of changing grain charge with $\Ubar_{0}=1$ (solid line), $\Ubar_{0}=0.3$ (dashed line), $\Ubar_{0}=3$ (dot-dashed line). Lines terminate at the small and large scales given by \cref{eq: WIM damping scale range}.
For the turbulent damping, the shaded region shows the range for $l_{\rm turb}$ in the range $100{\rm pc}$ to $300{\rm pc}$ with 
$M_{A}\approx1$. For nonlinear Landau damping we indicate the damping rate for the energetically dominant $\sim\!{\rm GeV}$ CRs, which, because it is nonlinear, depends on the CR properties and transport model.  For basic comparison, we use $l_{{\rm CR}}=300{\rm pc}$ and $e_{{\rm CR}}$  and in the range $0.1{\rm eV}\,{\rm cm}^{-3}$ to $10{\rm eV}\,{\rm cm}^{-3}$ ($e_{{\rm CR}}\approx0.3{\rm eV}\,{\rm cm}^{-3}$ is measured around the solar circle).  \revchng{Although our 
focus is on CR transport in well-ionised gas, we also show ion-neutral damping rates with $f_{\rm ion}=0.9$ and $f_{\rm ion}=0.99$ in the left-hand  panel (green dashed lines) in order to allow for a basic
comparison.  }
}
\label{fig:WIM}
\end{center}
\end{figure*}

Here we ask the question of whether dust
damping, as derived in \cref{sec:aw.dust.interation}, can compete with turbulent processes and
nonlinear Landau damping to  damp parallel waves and enable the fast transport of CRs (see \cref{sec: CR stuff}).
This simply involves directly computing the damping rate from \cref{eq:growth.rate} to compare 
with these previously studied processes. 
We compare our expressions graphically, for a range of reasonable warm-ISM parameters, in  \cref{fig:WIM}.

\subsubsection{Dust-induced damping}\label{subsub: dust damping WIM}

In the bulk of the ISM, radiation-pressure forces are relatively weak and grain drift is expected to be significantly sub-Alfv\'enic (\citealp{Weingartner2001a}; see figure 6 of \hsmag). It is thus reasonable to set $w_{s}=0$, or $\mach=-1$
in evaluating \cref{eq:growth.rate}, which also implies that $t_{s}$ does not directly enter the analysis (although we 
must ensure $t_{s}\omega_{A}\gg1$ for the modes of interest for \cref{eq:growth.rate} to be valid). 
For concreteness, we start by considering  grain charging in the absence of a radiation field (\cref{eq: grain charge no photoelectric}), which gives the Larmor time (\cref{eq: tL definition})
\begin{equation}
t_{L}\approx -1.7\times 10^{9}{\rm s} \frac{\adn{-5}^{2}\bar{\rho}_{d;0}}{\Ubar T_{4} B_{-6}}.
\end{equation}
Dust will damp CR-induced AWs so long as the scales that are resonant with CRs are also resonant with grains 
present in the ISM. The former resonant-CR scale, $k_{\rm res}^{-1}\approx r_{L}$, is given by \cref{eq: rL}, while the 
 dust-resonant wavenumber $k^{-1}=v_{A}|t_{L}|$ is 
 \begin{equation}
k^{-1}(\ad) = 3.3\times 10^{12}{\rm cm}\begin{cases}
\left(\frac{\ad}{8.2{\rm nm}}\right)^{2}\frac{\bar{\rho}_{d;0}}{n_{0}^{1/2}T_{4}\Ubar_{0}}& k\lesssim k_{\rm efe}\\
\left(\frac{\ad}{31{\rm nm}}\right)\frac{\bar{\rho}_{d;0}}{n_{0}^{1/2}\Ubar_{0}}& k\gtrsim k_{\rm efe}
\end{cases},\label{eq: resonant grain scale}
\end{equation}
where we have normalized the grain size to illustrate the connection to scales relevant for $R^{{\rm CR}}\approx 1{\rm GV}$ protons (see \cref{eq: rL}).
Here $k_{\rm efe}^{-1}\approx(2.2\times 10^{11}{\rm cm})\,{T_{4} \bar{\rho}_{d;0}}/(n_{0}^{1/2}\Ubar_{0})$ is the resonant scale at which the
grain charging changes from the collisional regime (large scales; $\adn{-5}\gtrsim T_{4}/47$) to the electron-field-emission-limited regime  (small scales; $\adn{-5}\lesssim T_{4}/47$) (see \cref{eq: grain charge no photoelectric}).
We also see from \cref{eq: xi tL xi w defs} that $\xi_{t_{L}}\approx2$ for $\adn{-5}\gtrsim T_{4}/47$ (larger scales), and  $\xi_{t_{L}}\approx 1$ for $\adn{-5}\lesssim T_{4}/47$ (smaller scales), showing that the damping rate is somewhat larger in the electron-field-emission-limited regime.\footnote{This occurs 
because the Larmor time increases more slowly with grain size, meaning there are more grains that are resonant with a particular wavelength of wave.} 
It is straightforward to verify that $t_{s}\omega_{A}\gg1$ for all grains (Coulomb drag dominates; \cref{eq: coulomb drag}).

The growth rate is computed by inverting \cref{eq: resonant grain scale} for $\ad$ and using the normalized power-law grain  mass distribution $d\bar{\mu}/d\ln \ad=\xi_{\mu}\ad^{\xi_{\mu}}/(\amax^{\xi_{\mu}}-\amin^{\xi_{\mu}})$. 
 \Cref{eq:growth.rate} then gives
\begin{align}
& \frac{\Gamma_{{\rm dust}}}{\omega_{A}}\approx \frac{\pi}{4}\frac{\mu_{0}}{\Delta^{\xi_{\mu}}_{\ad}}  \begin{cases} \left(\frac{k^{-1}}{4.9\times 10^{14}{\rm cm}}\right)^{\xi_{\mu}/2}\left(\frac{n_{0}^{1/2}T_{4}\Ubar_{0}}{\bar{\rho}_{d;0}}\right)^{\xi_{\mu}/2} & k_{\amax}\lesssim k\lesssim k_{\rm efe}\\
2 \left(\frac{k^{-1}}{1.0\times 10^{13}{\rm cm}}\right)^{\xi_{\mu}} \left(\frac{n_{0}^{1/2}\Ubar_{0}}{\bar{\rho}_{d;0}}\right)^{\xi_{\mu}} &   k_{\amin} \gtrsim k\gtrsim k_{\rm efe}
\end{cases}\nonumber \\
& \approx 
\begin{cases}
7.6\times 10^{-4} \left(\frac{\mu_{0}}{0.01}\right)\left(\frac{k^{-1}}{3.3\times 10^{12}{\rm cm}}\right)^{1/4}\frac{n_{0}^{1/8}T_{4}^{1/4}\Ubar_{0}^{1/4}}{\bar{\rho}_{d;0}^{1/4}} & k_{\amax}\lesssim k\lesssim k_{\rm efe}\\
3.0\times 10^{-3}\left(\frac{\mu_{0}}{0.01}\right)\left(\frac{k^{-1}}{3.3\times 10^{12}{\rm cm}}\right)^{1/2}\frac{n_{0}^{1/4}\Ubar_{0}^{1/2}}{\bar{\rho}_{d;0}^{1/2}} & k_{\amin} \gtrsim k\gtrsim k_{\rm efe}
\end{cases}\nonumber \\
& \approx 
\begin{cases}
6.5\times 10^{-4} \left(\frac{\mu_{0}}{0.01}\right)\left(\frac{R^{{\rm CR}}}{{\rm GV}}\right)^{1/4}\frac{T_{4}^{1/8}(\beta/10)^{1/8}\Ubar_{0}^{1/4}}{\bar{\rho}_{d;0}^{1/4}} & k_{\amax}\lesssim k\lesssim k_{\rm efe}\\
2.2\times 10^{-3}\left(\frac{\mu_{0}}{0.01}\right)\left(\frac{R^{{\rm CR}}}{{\rm GV}}\right)^{1/2}\frac{(\beta/10)^{1/4}\Ubar_{0}^{1/2}}{T_{4}^{1/4}\bar{\rho}_{d;0}^{1/2}} & k_{\amin} \gtrsim k\gtrsim k_{\rm efe}
\end{cases}
\label{eq: WIM damping}
\end{align}
where $\Delta^{\xi_{\mu}}_{\ad} = [(\amax/0.1\mu{\rm m})^{\xi_{\mu}}-(\amin/0.1\mu{\rm m})^{\xi_{\mu}}]/\xi_{\mu}\approx 3.0$
for grains in the range $\amin\approx1{\rm nm}$ to $\amax\approx0.25\mu{\rm m}$ with $\xi_{\mu}=0.5$, which 
is used to simplify the expressions from the second line. \Cref{eq: rL} is used to convert between scale and CR rigidity for the third line (assuming 
relativistic CRs, $R^{\rm CR}\gtrsim 1{\rm GV}$).
The wavenumber cutoffs, $k_{\amax}$ and $k_{\amin}$, are the wavenumbers resonant with the largest and smallest grains, 
respectively. Renormalizing $\ad$ in \cref{eq: resonant grain scale}, we find
\begin{gather}
k_{\amin}^{-1}\approx 1.0 \times 10^{11}{\rm cm}\frac{\bar{\rho}_{d;0}}{n_{0}^{1/2}\Ubar_{0}}\left(\frac{\amin}{1{\rm nm}}\right),\\ 
k_{\amax}^{-1}\approx3.1\times 10^{15}{\rm cm}\frac{\bar{\rho}_{d;0}}{T_{4}n_{0}^{1/2}\Ubar_{0}}\left(\frac{\amax}{0.25\mu{\rm m}}\right)^{2},
\label{eq: WIM damping scale range}\end{gather}
assuming that $\ad=\amin$ grains are in the field-emission-limited regime, and $\ad=\amax$ grains are in the collisional regime. 
We see that the resonant scales of grains, where there is strong damping, covers a similar range to that of $\sim \!{\rm GeV}$ CRs
across a wide range of parameters relevant to the well-ionized ISM.
Note that  \cref{eq: WIM damping} is discontinuous across $k\approx k_{{\rm efe}}$ (larger at small scales), because $\xi_{t_{L}}$ changes. 

In many regions of the ISM, while radiation pressures may not be sufficiently large 
to drive grains to super-Alfv\'enic velocities, the radiation field can nonetheless be strong enough for photoelectric
charging to dominate over collisional charging  \citep[e.g.,][]{Weingartner2001a,Draine2010}. In this 
case grains become positively charged (\cref{eq: grain charge photoelectric}) to several volts depending on the strength
of the radiation field. Aside from the polarization of the damped waves, which makes no difference to the
CR transport within our approximations (see \cref{subsec: CR polarization}), the calculation of the wave-damping rate 
is nearly identical except that there is no electron-field-emission limited regime. Taking $\Ubar=-\Ubar_{0}/T_{4}$ (corresponding to $U_{d}\approx 2.2{\rm V}\, \Ubar_{0}$), with 
$\Ubar_{0}$ encapsulating our ignorance about the radiation field, the same calculation as above gives the 
dust-resonant wavenumber
\begin{equation}
k^{-1}(\ad) = 3.3\times 10^{12}{\rm cm}\left(\frac{\ad}{8.2{\rm nm}}\right)^{2}\frac{\bar{\rho}_{d;0}}{n_{0}^{1/2}\Ubar_{0}},
\end{equation}
with damping rate
\begin{align}
\frac{\Gamma_{\rm dust}}{\omega_{A}} &\approx 7.6\times 10^{-4} \left(\frac{\mu_{0}}{0.01}\right)\left(\frac{k^{-1}}{3.3\times 10^{12}{\rm cm}}\right)^{1/4}\frac{n_{0}^{1/8}\Ubar_{0}^{1/4}}{\bar{\rho}_{d;0}^{1/4}}\nonumber\\
&\approx6.5\times 10^{-4} \left(\frac{\mu_{0}}{0.01}\right)\left(\frac{R^{{\rm CR}}}{{\rm GV}}\right)^{1/4}\frac{(\beta/10)^{1/8}\Ubar_{0}^{1/4}}{T_{4}^{1/8}\bar{\rho}_{d;0}^{1/4}}, \label{eq: WIM damping photoelectric}
\end{align}
between the minimum and maximum scales,
\begin{gather}
k_{\amin}^{-1}\approx 4.9 \times 10^{10}{\rm cm}\frac{\bar{\rho}_{d;0}}{n_{0}^{1/2}\Ubar_{0}}\left(\frac{\amin}{1{\rm nm}}\right)^{2},\\ 
k_{\amax}^{-1}\approx3.1\times 10^{15}{\rm cm}\frac{\bar{\rho}_{d;0}}{n_{0}^{1/2}\Ubar_{0}}\left(\frac{\amax}{0.25\mu{\rm m}}\right)^{2},
\end{gather}
which is nearly identical to the collisionally charged result above.
Of course, it is  plausible that in some regions photo-electric and collisional charging will cancel out, in which 
case $|\Ubar|$ could be reduced substantially compared to the estimates made above; however, it is reasonable to expect that 
in most regions  one of the two charging regimes should dominate over the other.

\subsubsection{Other damping mechanisms}\label{subsub: other damping WIM}

\Cref{eq: WIM damping} should be compared to the damping
rate of parallel AWs from other processes. The most important is turbulent damping, which
arises from the fact that parallel AWs excited by CRs must propagate along  inhomogenous turbulent
magnetic fields.  Using the \citet{Goldreich1995} phenomenology of anisotropic magnetized turbulence,
the damping rate of $r_{L}\approx k^{-1}$ scale waves is \citep{Farmer2004,Lazarian2016,Zweibel2017}
\begin{align}
\frac{\Gamma_{\rm turb}}{\omega_{A}}& \approx \frac{1}{\omega_{A}}\frac{v_{A}}{r_{L}^{1/2}l_{A}^{1/2}}\nonumber\\
\approx &1.0\times 10^{-4}\,\Lambda_{\beta} \left( \frac{k^{-1}}{3.3\times 10^{12}{\rm cm}} \right)^{1/2} M_{A}^{1/2\zeta}\left(\frac{l_{\rm turb}}{100{\rm pc}}\right)^{-1/2}\nonumber\\
\approx & 7.6\times 10^{-5}\,\Lambda_{\beta} \left( \frac{R^{{\rm CR}}}{{\rm GV}} \right)^{1/2} \left( \frac{\beta/10}{n_{0}T_{4}}\right)^{1/4}M_{A}^{1/2\zeta}\left(\frac{l_{\rm turb}}{100{\rm pc}}\right)^{-1/2}. \label{eq: turbulent damping}
\end{align}
Here $l_{\rm turb}$ is the outer scale of the turbulence and $l_{A}=M_{A}^{-1/\zeta}l_{\rm turb}$ is the scale at which turbulent fluctuations become sub-Alfv\'enic, with $M_{A}$  the Alfv\'en Mach number at $l_{{\rm turb}}$ and $\delta u\sim l^{-\zeta}$  the power-law turbulent scaling exponent of $l>l_{A}$ motions ($\zeta\sim1/2$ for supersonic motions, or $\zeta\sim1/3$ for subsonic motions).
The factor  $\Lambda_{\beta}\approx\max(1,0.4\!\sqrt{\beta})$  arises because at lower $\beta$, direct turbulent dissipation of perpendicular waves is expected to dominate \citep{Farmer2004}, while at larger $\beta$, linear-Landau damping of perpendicular magnetosonic waves dominates \citep{Zweibel2017,Wiener2018}. \revchng{Although 
Alfv\'enic turbulence is known to be quite robust across a wide range of plasma conditions, it is worth noting that any
other processes that enhance the damping of turbulence beyond these standard estimates (e.g., \citealp{Silsbee2020}) 
would reduce small-scale fluctuations in the magnetic field, thus decreasing $\Gamma_{\rm turb}$ and the importance of turbulence to CR propagation.}

Nonlinear Landau damping (NLLD)  is the process by which nonlinear magnetic-field strength variations in 
the small-scale parallel AWs are directly damped by resonant particle interactions  \citep{Lee1973,Cesarsky1981,Voelk1982}
and pressure anisotropy \citep{Squire2016,Squire2017}. As a nonlinear effect, the strength of the damping depends on 
the amplitude of the waves as 
\begin{equation}
{\Gamma_{\rm NLL}}\approx\frac{ \sqrt{\pi}}{8} v_{\rm th} k \frac{\delta \bm{B}^{2}(k)}{B^{2}},
\end{equation}
where the wave amplitude ${\delta \bm{B}^{2}(k)}/{B^{2}}$
itself depends sensitively on the CR energy density and its gradient. 
In order to make a basic comparison to other mechanisms, we use convenient expressions
from \hcr\ (equation A4) for the damping rate of waves resonant with $\sim\!{\rm GeV}$ CRs. These are derived using an approximate balance of growth and damping, assuming that the waves
are excited by CRs with energy density $e_{{\rm CR}}$ that varies over length scale $l^{{\rm CR}}$ (see also \citealt{Thomas2019}), giving
\begin{align}
\frac{\Gamma_{{\rm NLL}}}{\omega_{A}}&\approx \frac{1}{\omega_{A}}\left[ \frac{1}{3}\frac{\pi^{1/2}}{8}\left(\frac{c_{s}v_{A}}{r_{L}l_{{\rm CR}}}\right) \left(\frac{e_{{\rm CR}}}{B^{2}/8\pi}\right)\right]^{1/2}\nonumber\\
&\approx 1.4\times 10^{-4} \left( \frac{k^{-1}}{3.3\times 10^{12}{\rm cm}} \right)^{1/2} \frac{(\beta/10)^{3/4}}{n_{0}^{1/2}T_{4}^{1/2}} \nonumber \\
&\qquad\qquad\times\left( \frac{e_{\rm CR}}{1{\rm eV}\,{\rm cm}^{-3}}\right)^{1/2}\left(\frac{l_{\rm CR}}{100{\rm pc}}\right)^{-1/2}
\label{eq: NLL damping}\end{align}
However, estimating the rigidity dependence of this damping rate requires a rigidity-dependent model of
CR transport (which manifests in \cref{eq: NLL damping} through the rigidity dependence of $e_{\rm CR}$). While we suggest a way to 
estimate this below (\cref{subsub: implications and transport}), leading to the scaling $\Gamma_{\rm NLL}/\omega_{A}\propto (R^{\rm CR})^{0.15}$, given the greater uncertainty in these estimates, in \cref{fig:WIM} 
we plot only the damping rate for $\sim\!{\rm GeV}$ CRs.
In any case,  \cref{eq: NLL damping} and the results of \hcr\ suggest that NLLD is subdominant and unimportant in most situations, being overwhelmed by turbulent damping even in the absence of dust.

Finally, although we do not consider partially ionized gas and ion-neutral 
damping in detail, it is helpful to include for comparison purposes. For waves with frequencies below the ion-neutral 
collision rate $\tnu_{{\rm in}}$, the AW damping rate is  (\hcr; \citealp{Amato2018})
\begin{align}
\frac{\Gamma_{{\rm in}}}{\omega_{A}}&\approx \frac{\tnu_{{\rm in}}}{\omega_{A}}\approx 0.04 \left( \frac{k^{-1}}{3.3\times 10^{12}{\rm cm}}\right)(1-f_{{\rm ion}})f_{\rm ion}^{1/2}\frac{T_{4}^{1/2}\rho_{-24}^{3/2}}{B_{-6}},\label{eq: in damping}
\end{align}
where $\rho_{-24}=\rho/(10^{-24}{\rm g}\,{\rm cm}^{-3})$ (mass density of both ions and neutrals), and the $f_{\rm ion}^{1/2}$ factor arises because $v_{A}$ is a function of the  ion density alone at small scales in a partially ionized gas (note that we have neglected this effect
in \crefrange{eq: turbulent damping}{eq: NLL damping} because it is unimportant if $1-f_{\rm ion}\ll1$). It is clear that ion-neutral damping dominates over all others 
when the neutral fraction $1-f_{\rm ion}$ is modestly large.
\begin{figure}
\begin{center}
\includegraphics[width=1.0\columnwidth]{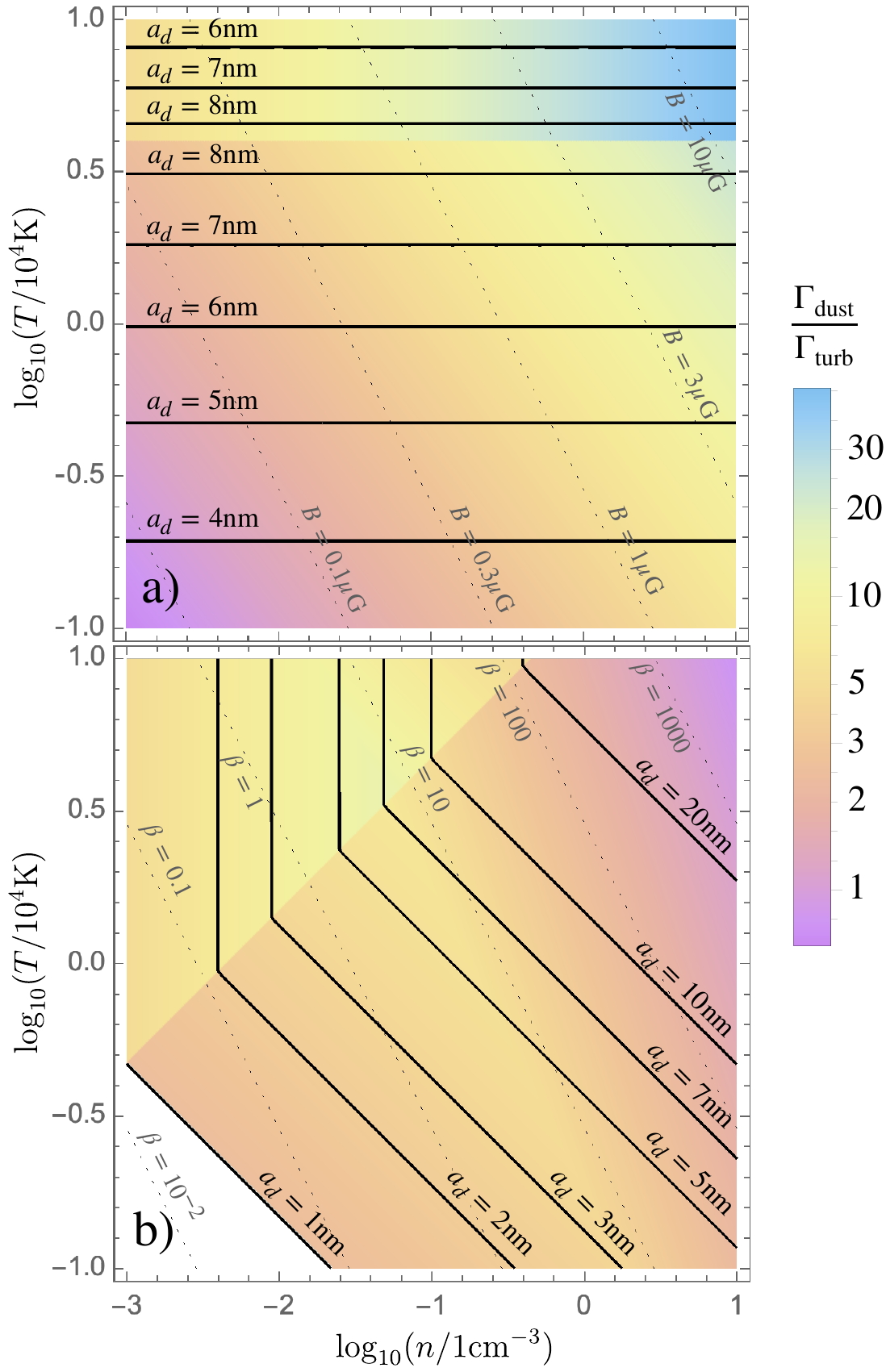}
\caption{Colours illustrate the ratio $\Gamma_{\rm dust}/\Gamma_{\rm turb}$ of the dust damping rate to turbulent damping rate for
scales corresponding to $R^{\rm CR}=1{\rm GV}$ CRs, across a wide range of temperatures and 
densities relevant to the ionized ISM. Collisional charging (\cref{eq: WIM damping}) is assumed for $\Gamma_{\rm dust}$.
The top panel keeps $\beta=10$ constant as $n$ and $T$ are changed, while the bottom panel 
keeps $B=1\mu{\rm G}$ constant. In each panel, the solid contours show the size of the resonant grains (\cref{eq: resonant grain scale}) as labeled,
while the light, dashed contours show lines of constant $B$ (top panel) or constant $\beta$ (bottom panel).
\revchng{The step change in color in each panel in each panel ($T_{4}\approx 10^{0.6}$ in the top panel; the  diagonal from $T_{4}\approx 10^{-0.3}$ in the bottom panel) is where }the charging becomes electron-field-emission limited, and the white region
in the bottom panel occurs because the resonant scale moves below $a=1{\rm nm}$. Other relevant parameters
are the same as \cref{fig:WIM}, with $\Ubar_{0}=1$ and $l_{\rm turb}=100{\rm pc}.$  }
\label{fig: regions GeV}
\end{center}
\end{figure}

\subsubsection{Implications for cosmic-ray transport }\label{subsub: implications and transport}

A graphical comparison of  \crefrange{eq: WIM damping}{eq: in damping}  in two sets of conditions relevant to 
the warm ISM, across a wide range of scales, is shown in  \cref{fig:WIM}. 
A clear conclusion is that the dust damping rate of AWs with a $\sim\!1\%$ dust to gas mass ratio is significantly larger than 
any other source of damping for AWs resonant with $\sim\!{\rm GeV}$ CRs (top axes), except ion-neutral damping  with 
$1-f_{\rm ion}\gtrsim 0.03$. In other words, \emph{dust may significantly decrease CR confinement in well-ionized regions of the ISM}.
The same conclusion is illustrated across a wider range of gas conditions in \cref{fig: regions GeV}, which  shows the ratio of
dust to turbulent damping
at the ${\rm GeV}$ CR gyroscale, with solid contours 
illustrating the resonant grain size (assuming collisional charging).
We see that it is generally very small grains ($\ad\lesssim 10{\rm nm}$ for common ionized-ISM conditions) that are resonant with 
scales relevant for $\sim\!{\rm GeV}$ CRs. 

Assuming that dust damping does dominate as suggested by the estimates above, we can balance the dust damping rate \eqref{eq: WIM damping} to the streaming instability growth rate to 
obtain an estimate for the effective speed of CR energy transport, $\bar{v}_{\rm st} $, in the presence of dust  (see, e.g., \hcr\ section 3.3).
For collisionally charged grains (\cref{eq: WIM damping}), one finds
\begin{align}
\bar{v}_{\rm st} &\approx v_{A} + \frac{3}{4} \frac{\kappa_{\|}}{l_{\rm CR}}\approx v_{A} + \frac{4c r_{L}}{\pi}\frac{\Gamma_{\rm damp}}{v_{A}} \frac{B^{2}}{8\pi }\frac{1}{e_{\rm CR}}\nonumber \\
\approx & v_{A} 
\begin{cases}
1+ 4.1 \left(\frac{R^{\rm CR}}{1{\rm GV}} \right)^{1/4}\left( \frac{e_{\rm CR}}{1{\rm eV}\,{\rm cm}^{-3}}\right)^{-1}\left(\frac{\mu_{0}}{0.01}\right)  \frac{n_{0} T_{4}^{5/8} \Ubar_{0}^{1/4}}{(\beta/10)^{3/8}\bar{\rho}_{d;0}^{1/4}} & R^{\rm CR}\gtrsim R^{\rm CR}_{\rm efe} \\
1+ 14 \left(\frac{R^{\rm CR}}{1{\rm GV}} \right)^{1/2}\left( \frac{e_{\rm CR}}{1{\rm eV}\,{\rm cm}^{-3}}\right)^{-1}\left(\frac{\mu_{0}}{0.01}\right)  \frac{n_{0} T_{4}^{1/4} \Ubar_{0}^{1/2}}{(\beta/10)^{1/4}\bar{\rho}_{d;0}^{1/2}} & R^{\rm CR}\lesssim R^{\rm CR}_{\rm efe}\end{cases}\label{eq:vst estimate collisional}
\end{align}
where $\kappa_{\|}=c^{2}/3\tnu_{c}$ is the parallel CR diffusion coefficient and  $R^{\rm CR}_{\rm efe}\approx 0.4 T_{4}^{3/2}\bar{\rho}_{d;0}/(\Ubar_{0}\beta^{1/2}){\rm GV}$ is the 
CR rigidity corresponding to $k_{\rm efe}$. Similarly, for 
photoelectrically charged grains (\cref{eq: WIM damping photoelectric}) one finds
\begin{equation}
\bar{v}_{\rm st} \approx v_{A} \left[1+ 4.1 \left(\frac{R^{\rm CR}}{1{\rm GV}} \right)^{1/4}\left( \frac{e_{\rm CR}}{1{\rm eV}\,{\rm cm}^{-3}}\right)^{-1}\left(\frac{\mu_{0}}{0.01}\right)  \frac{n_{0} T_{4}^{3/8} \Ubar_{0}^{1/4}}{(\beta/10)^{3/8}\bar{\rho}_{d;0}^{1/4}}. \right]\label{eq:vst estimate pe}
\end{equation}
 Note
that \cref{eq:vst estimate collisional,eq:vst estimate pe} apply only to the $\sim\!{\rm GeV}$ CRs that dominate the
CR energy density, because of the appearance of $e_{\rm CR}$ in the expressions, and we also assume relativistic CRs to convert scales to rigidity using \cref{eq: rL}.
Evidently, we predict modestly super-Alfv\'enic CR energy transport with faster transport at higher temperature, lower $\beta$, or with more strongly 
charged grains. It is, however, worth noting that this streaming speed likely remains too small to 
fully explain the discrepancy found by \hcr\ between observations and cosmological simulations with self-confined CRs
(or equivalently, the AW damping rate 
remains too small). Further simulations with a dust-damping self-confinement model would be needed to test this more thoroughly.

Finally, we can estimate the predicted rigidity dependence of self-confined  CR transport regulated by dust, $\tilde{v}_{\rm st}\propto (R^{\rm CR})^{\delta}$ (or equivalently $\kappa_{\|}\propto (R^{\rm CR})^{\delta}$). Such predictions 
can be compared to the measured rigidity dependence of CR grammage from, e.g., the B/C ratio, which 
suggests $\delta$ in the range $0.3$ to $0.8$ (with various uncertainties related to sources and propagation; \citealp{Maurin2010,Amato2014,Blasi2017,Aguilar2018}). To estimate $\delta$, we follow the kinetic treatment of
\cite{Skilling1971},
which, by balancing the damping rate of AWs to the growth rate of the streaming instability, arrives
at the the following equation governing the distribution $f(p)$ of CRs of momentum $p$:
\begin{equation}
\left(\frac{\partial}{\partial t}+\bm{w}\cdot\nabla\right) f=\frac{1}{3}p\frac{\partial f}{\partial p}\nabla\cdot\bm{w} - \frac{1}{p^{3}}\nabla\cdot\left( \frac{c\Gamma_{\rm damp}(p) B}{4\pi^{3}e v_{A}}\hat{\bm{b}}\right),\label{eq: skilling}
\end{equation}
where $\bm{w}=\bm{u}+v_{A}\hat{\bm{b}}$ is the velocity of the AW frame, in which \cref{eq: skilling} is valid. 
The wave-damping rate $\Gamma_{\rm damp}$ has an effective momentum dependence\footnote{\cite{Skilling1971} neglects the $k$ dependence of $\Gamma_{\rm damp}$ in his calculation. Including it yields slightly different  (unimportant) numerical factors in the final term of \cref{eq: skilling}.} because it is in general a function
of $k$, and $k$ and $p$ are related by the CR resonance condition \cref{eq: rL}.  We see that 
the final ``damping'' term on the right-hand side, which necessarily describes any deviation from 
transport at exactly $\tilde{v}_{\rm st}=v_{A}$ within these approximations, does not have the form of an 
advection or diffusion operator at all, and in fact does not even contain the CR distribution $f$ (its rather 
bizarre behavior is discussed in \citealt{Skilling1971}). Nonetheless, as a very 
simple, approximate  estimate for the scaling of an effective transport speed $\tilde{v}_{\rm st}$ with CR momentum, 
we can compare this damping term to an advection term, which would have the form $\tilde{v}_{\rm st}\hat{\bm{b}}\cdot\nabla f$.
Noting that $f\sim p^{-\alpha}$ with $\alpha\approx4.7$, while $\Gamma_{\rm damp}$ scales as $\Gamma_{\rm damp}\sim k^{b}\sim p^{-b}$, we can estimate $\delta$ by matching the $p$ scaling of  $\tilde{v}_{\rm st}\hat{\bm{b}}\cdot\nabla f$ to that of 
$p^{-3}\Gamma_{\rm damp}$, giving $\delta \approx \alpha-3-b=1.7-b$. From
\cref{eq: WIM damping}, we see $b=1-\xi_{\mu}/2\approx 0.75$, giving $\delta \approx 0.7+\xi_{\mu}/2\approx 0.95$. Although this estimate is a little higher than observationally favoured values, it is worth noting
that the other AW damping mechanisms except nonlinear-Landau damping\footnote{The scaling of 
NLLD can be worked through in the same way, with the caveat that $\Gamma_{\rm damp}$ depends directly 
on the CR spectrum in Skilling's argument. This leads to a damping term that scales as $\sim p^{-3/2}\nabla\cdot[A (\hat{\bm{b}}\cdot\nabla f)^{1/2}\hat{\bm{b}}]$ (where $A$ represents terms that depend on gas quantities), which gives $\delta=0.85$. This
also implies that $\Gamma_{\rm NLL}$ from \cref{eq: NLL damping} scales as $\Gamma_{\rm NLL}\sim k^{0.85}$ for
this model of CR transport. For comparison, this is slightly flatter than the other mechanisms plotted in \cref{fig:WIM}. This calculation is effectively the same as that of \cite{Blasi2019}, who also finds $\delta =0.85$ for NLLD-mediated transport.
} all predict larger $\delta$ (e.g., $\delta \approx 1.2$ for
turbulence damping). A flatter (or even inverted) mass distribution of small grains, as favoured by more detailed models of dust (see \cref{subsub:mass distribution}), would decrease $\delta$ further.

\subsubsection{Uncertainties}

The most significant uncertainties in the above conclusions relate to the fact that a reasonable population of small grains  ($\ad\lesssim 10{\rm nm}$)
is  required to  damp waves relevant to $\sim\!{\rm GeV}$ CRs. 
If the total dust-to-gas-mass ratio were significantly lower than $0.1\rightarrow 0.3\%$ in more diffuse ISM regions \citep[see, e.g.,][]{Peters2017},
or if the grain size distribution 
were significantly depleted for $\ad\lesssim 10{\rm nm}$, dust damping could be unimportant compared to other sources.
However, it is worth noticing that most complex dust models that are calibrated to observations \citep[e.g.,][]{Weingartner2001,Zubko2004,Draine2009} involve a grain distribution with an \emph{excess} of the smallest grains  compared to a power-law distribution, which suggests that using  $\xi_{\mu}=0.5$ could underestimate the dust damping,
as well as overestimating the dependence of transport on CR rigidity ($\delta$).
Further uncertainty concerns the charging of  small grains, which will  be affected by charge quantization  \citep{Draine2010} as well as 
the environmental factors discussed above (e.g., the radiation environment); 
a more accurate treatment could convolve a charge distribution with the mass distribution 
before computing the damping rate to account for  quantization.
Finally, we reiterate that only one wave polarization is damped by dust (see \cref{subsub: polarization}),
so that in practice, dust tends to make CR-induced AWs more circularly polarized.
As discussed in \cref{subsec: CR polarization}, this likely enhances CR transport by a similar degree to what would occur
if both polarizations were damped; however, simulations and more accurate theory are clearly needed 
to assess this physics in more detail.

\subsection{Enhanced cosmic-ray scattering through dust-excited Alfv\'en waves}\label{sub: CGM}

In this section, we discuss the possibility that dust drifting with $|\driftvel |>v_{A}$  generates  small-scale parallel Alfv\'en waves,
which subsequently scatter and confine cosmic rays. Much of this physics remains highly uncertain; both the astrophysics of dust in the relevant conditions
(e.g., charging and grain abundances in hotter gas) and  the nonlinear wave physics (e.g., saturation of the dust instability and CR scattering; see \cref{subsec: CR polarization}) are not well constrained. We thus provide only a cursory examination of the relevant physics compared to \cref{sub: WIM} on dust 
damping. Nonetheless, in some situations -- particularly when grains are driven to significantly super-Alfv\'enic velocities by 
radiation pressure -- it is reasonable to suggest that dust-induced scattering rates could dominate over all other CR scattering mechanisms, causing
scattering levels near the ``Bohm limit,'' where the CR's mean-free path  approaches their gyroradius.
Here, motivated by \hsmag, we consider the interesting example of the circumgalactic medium (CGM) around a  quasar or highly luminous galaxy, which
could cause an interesting nonlinear correlation between the luminosity and the confinement of CRs.
Other
possible applications include near supernovae\footnote{Of course, CR  fluxes are also particularly extreme near supernovae, and 
the relevant processes remain highly uncertain \citep{Micelotta2018,Bykov2018,Holcomb2019}.} and regions of the ISM with higher-than-average radiation fields   (see, e.g., figure 6 and section 9 of \hsmag), although the dust instabilities may become more complicated in some cases where grains are not so strongly magnetized (if the ratio $|t_{L}|/t_{s}$ approaches one, see \cref{sub: aw dust setup}).

As shown in \cref{sec:aw.dust.interation}, positively charged dust drifting at speeds exceeding $v_{A}$ will generate 
 parallel Alfv\'en waves. If their growth rate exceeds the damping rate from background turbulence,\footnote{Nonlinear Landau damping will 
presumably not operate on the dust-excited AWs, since they do not involve a variation in $|\bm{B}|$.} such waves will presumably 
continue to grow until they saturate through some other means. The most obvious candidate is by exciting 
some type of turbulence, which scatters dust particles sufficiently to shut off the instability's drive. The simulations of \cite{Hopkins2020} suggest that a simple saturation criterion is when a characteristic turnover time 
at the scale of interest is equal to the instability growth rate, which leads to an estimate of the power in small scale parallel fluctuations $\delta \bm{B}^{2}/B^{2}\sim (\Gamma/\omega_{A})^{2}$, where $\Gamma/\omega_{A}$ is the growth rate of the dust instability normalized by the Alfv\'en frequency at the chosen scale.
This implies the CR diffusivity 
\begin{equation}
\kappa_{\|}\approx \frac{c^{2}}{3\tnu_{c}}\sim 3.3 \times 10^{26}{\rm cm}^{2}{\rm s}^{-1}\left(\frac{ R^{{\rm CR}}}{{\rm GV}}\right)\left(\frac{B}{1\mu{\rm G}}\right)^{-1}\left(\frac{\Gamma/\omega_{A}}{0.01}\right)^{-2}.\label{eq: kappa from dust modes}
\end{equation}
It is worth noting that  this estimate $\delta \bm{B}^{2}/B^{2}\sim (\Gamma/\omega_{A})^{2}$ likely significantly underestimates the saturation level of small-scale AWs,
because the instability grows with circularly polarized modes that do not perturb the field strength, so will 
not necessarily break up into turbulent fluctuations in the usual way. On the other hand, it is 
also likely the  case that  pure circularly polarized modes are very inefficient at isotropizing CRs (see \cref{subsec: CR polarization}), so 
\cref{eq: kappa from dust modes} could be a reasonable estimate of effective scatterers (i.e., turbulent fluctuations of both polarizations) even 
if one polarization reaches higher amplitude.   Unfortunately,
 the extremely small scales of interest were not directly probed by the simulations of \citealt{Hopkins2020} (the ``CGM'' simulation approaches similar conditions). Further simulations or more detailed analytic theory are clearly needed, but we proceed anyway for lack of a more accurate estimate.

\begin{figure}
\begin{center}
\includegraphics[width=1.0\columnwidth]{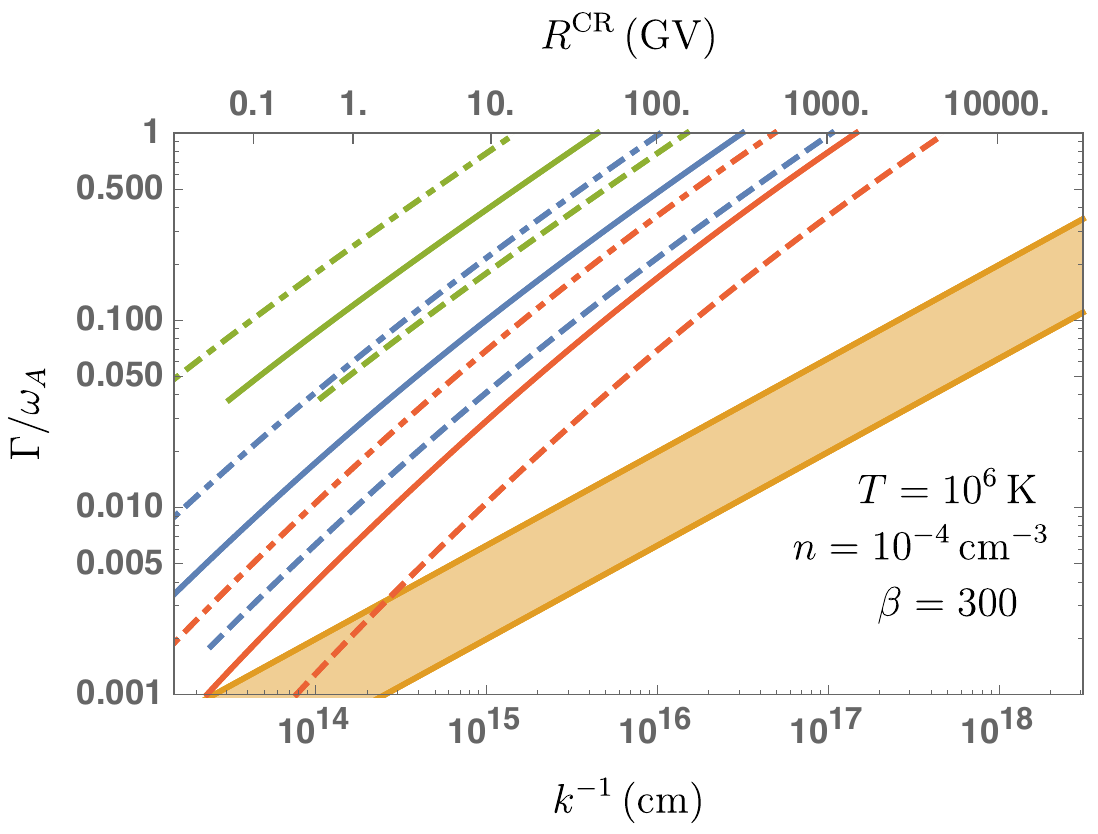}
\caption{The {growth rate} of unstable AWs in the presence of drifting dust (lines) compared to  turbulent damping rates (orange shaded region), for conditions relevant to the CGM around a luminous quasar ($T=10^{6}{\rm K}$, $n=10^{-4}{\rm cm}^{-3}$, $\beta=300$). We take $\mu_{0}=0.01$
with photoelectrically charged grains and an MRN grain spectrum between $\amin=1{\rm nm}$ and $\amax=0.25\mu{\rm m}$. The blue, green, and red lines respectively show growth rates with an $L\approx 10^{13}L_{\sun}$,
$L\approx 5\times 10^{13}L_{\sun}$, and $L\approx 3\times10^{12}L_{\sun}$ source, all at $r\sim 100 {\rm kpc}$. As in \cref{fig:WIM}, solid, dot-dashed, and dashed lines respectively show $\Ubar=1$, $\Ubar=3$, and $\Ubar=0.3$, to illustrate the effect of changing the grain charge, and the lines terminate on the left 
at the scale where $\amin=1{\rm nm}$ grains are resonant (\cref{eq: kmin for CGM}). We take $\Phi=1$ since its scaling is degenerate with $L$ or $r$. For turbulent damping, we use the expression \eqref{eq: turbulent damping}, with the range indicating the a range of outer-scales from $l_{\rm turb}\sim 10{\rm kpc}$ to $l_{\rm turb}\sim 100{\rm kpc}$ with $M_{A}\sim 5$.
}
\label{fig:CGM}
\end{center}
\end{figure}

We now focus on conditions relevant to the CGM around a  quasar for concreteness, although 
similar considerations could apply to any high-luminosity galaxy. We first note that significant amounts of dust is observed in the CGM \citep{Menard2010,Peek2015}, presumably driven there by radiation pressure \citep[e.g.,][]{Ishibashi2015,Hirashita2019}. Considering conditions appropriate to distances  $r\sim r_{100}100{\rm kpc}$ from a quasar of luminosity $L\sim L_{13} 10^{13}L_{\sun}$, we take the gas to be hot ($T\sim 10^{5}\rightarrow 10^{7}{\rm K}$), at very low density ($n\sim 10^{-6}\rightarrow 10^{-3}{\rm cm}^{-3}$), and with weak magnetic fields ($\beta\sim 100\rightarrow 10^{4}$).\footnote{A  potential complication and uncertainty is that the observed dust may not
be cospatial with the hot gas in the CGM, which is known to be multiphase and clumpy \citep{Tumlinson2017}. In such 
a case, although our estimates could still apply to waves in the cooler, denser  gas (suggesting high CR scattering rates therein), isolated clumpy regions with high scattering rates are likely to be ineffective at confining CRs (since the CRs fill the space between clumps; \hcr). On
the other hand, most dust grains drift supersonically through the gas (\cref{eq: GCM drift erad}), so will presumably 
blow through different gas phases, even if initially present only in a cooler phase. Clearly, the relevance of multiphase gas  represents yet another
significant uncertainty for this mechanism.  } The first stage of estimating the AW growth-rate from \cref{eq:growth.rate} is to estimate the Larmor 
time and relative drift velocity of grains. Estimates of $\psi$ indicate that grains will be strongly photoelectrically charged (\cref{eq: grain charge no photoelectric}), 
suggesting $\Ubar\approx 3/T_{4}$ ($U_{d}\sim 7{\rm V}$) and that Epstein drag dominates  (particularly given
that we find trans-sonic to supersonic drift velocities; c.f.~\cref{eq: epstein drag,eq: coulomb drag}). We estimate the radiative force on the  
grains as $ m_{d}\accel\sim Q_{\rm abs} \pi \ad^{2} L/(4\pi r^{2}c)$, where $Q_{\rm abs}$ is the absorption efficiency of the grains, which, for a radiative flux peaked around wavelength $\lambda_{\rm rad}$,  is $Q_{\rm abs}\sim \ad/\lambda_{\rm rad}$ for $\ad\lesssim \lambda_{\rm rad}$ or $Q_{\rm abs}\sim 1$ for larger grains \citep{Weingartner2001b}. \revchng{Assuming, for simplicity, that the relative dust-gas drift is supersonic ($\ws \gg c_{s}$) to solve $\ws=\accel t_{s}^{\rm Ep}$ for $\ws$, we find}
\begin{equation}
\frac{\ws}{v_{A}}\approx \mach \approx 330 \,\adn{-5}^{1/2} \frac{L_{13}^{1/2}}{ r_{100}} {B_{-8}}^{-1} \Phi \label{eq: GCM drift erad}
\end{equation}
for dust with $\ad\lesssim\lambda_{\rm rad}$ and a spectrum that peaks around Lyman-alpha wavelengths, $\lambda_{\rm rad}=122{\rm nm}$ (we define $B_{-8}=B/10^{-8}{\rm G}$). \revchng{The projection factor $\Phi \equiv (\baccel\cdot\bm{B}_{0})/(|\baccel||\bm{B_{0}}|)$ accounts for a possible misalignment of the magnetic field with the
radiative flux that accelerates grains (see \cref{sub: aw dust setup}).} \Cref{eq: CGM growth rate} somewhat \revchng{overestimates $\ws$ for the smallest grains that are trans- or sub-sonic ($\ws/v_{A}\lesssim \beta^{1/2}$), as well as for the largest grains with $\ad\gtrsim\lambda_{\rm rad}$ ($Q_{\rm abs}$ saturates at $Q_{\rm abs}\sim 1$); these affects are correctly accounted for in \cref{fig:CGM}}. \Cref{eq: CGM growth rate} also does not allow for possible nonlinear modifications to $\ws$ as the grains are scattered by the turbulence they induce
(although this is likely a modest effect in this very short wavelength range; \citealp{Moseley2019}).
 The Larmor time is $t_{L}\approx (5.3\times 10^{10}{\rm s})\,\adn{-5}^{2} \bar{\rho}_{d;0}/(\Ubar_{0} B_{-8})$ using 
the photoelectric expression \eqref{eq: grain charge photoelectric} for the grain charge. Inverting the resonant condition  $k^{-1}=v_{A}t_{L}\mach$ to link the
grain size to wavelength, and assuming $\xi_{\mu}=0.5$ for $\amin\approx1{\rm nm}$ to $\amax\approx0.25\mu{\rm m}$, \cref{eq:growth.rate} becomes
\begin{align}
\frac{\Gamma}{\omega_{A}}\approx  0.13 \left(\frac{\mu_{0}}{100}\right)&\left(\frac{k^{-1}}{3.3\times 10^{14}{\rm cm}}\right)^{3/5}\!\frac{(\beta/300)^{7/10}}{n_{-4}^{2/5}T_{6}^{7/10}} \nonumber \\
&\times \left(\frac{L_{13}}{r_{100}^{2}}\right)^{7/10}\left(\frac{\Ubar_{0}}{\bar{\rho}_{d;0}}\right)^{3/5}\Phi^{7/5},\label{eq: CGM growth rate}
\end{align}
where we have normalized $k^{-1}$ to the scale resonant with $\sim\!{\rm GeV}$ particles at $B= 10^{-8}{\rm G}$. \Cref{eq: CGM growth rate} applies
on scales $k<k_{\amin}$ where resonant grains exist, with
\begin{equation}
k_{\amin}^{-1}\approx 1.5\times 10^{13}{\rm cm}\, \left(\frac{\amin}{1{\rm nm}}\right)^{5/2}\!\!\frac{(\beta/300)^{1/2}}{n_{-4}T_{6}^{1/2}}\frac{L_{13}^{1/2}}{r_{100}}\frac{\bar{\rho}_{d;0}}{\Ubar_{0}}\Phi.\label{eq: kmin for CGM}
\end{equation}
We see that growth rates are rather large at the 
relevant scales for $\sim\!{\rm GeV}$ CRs, and, at modestly larger scales relevant to  $\gtrsim\!50{\rm GeV}$ particles, the scattering level might be expected reach Bohm diffusion levels (this occurs when $\delta \bm{B}/B\sim1$ from $\Gamma\sim \omega_{A}$, although our solution for $\Gamma$ breaks down here also). However, as also occurred for dust damping in the ISM (\cref{sub: WIM}), it is the smaller grains that grow AWs at $\sim\!{\rm GeV}$ scales, and if the grain spectrum were cut off at small scales, or grains were less charged than expected, the spectrum of unstable AWs may not extend to sufficiently small scales to efficiently scatter $\sim\!{\rm GeV}$ particles.

We illustrate the AW growth rate for the fiducial conditions  ($T=10^{6}{\rm K}$, $n=10^{-4}{\rm cm}^{-3}$, $\beta=300$)  with $\mu_{0}=0.01$ in \cref{fig:CGM} (although $\mu_{0}=0.01$ may be an overestimate, the growth rate simply scales linearly with $\mu_{0}$). 
We use the full Epstein-drag expression   to compute $\ws$, without assuming $\ws\gtrsim c_{s}$ or $\ad\lesssim\lambda_{\rm rad}$ as in \cref{eq: GCM drift erad}, which implies
that the curves are slightly below the estimate \eqref{eq: CGM growth rate} across most scales. Overall, we see significant
dust-induced AW growth rates that are large compared to the effect of turbulence (shaded region), which
provides a basic measure of the expected growth rate of the CR streaming instability in such regions. A caveat is
that in order to scatter  $\sim\!{\rm GeV}$ particles, there must be a population of quite small grains. Comparing the estimate
of CR diffusivity in \cref{eq: kappa from dust modes} with the  estimate of $\kappa_{\|}\gtrsim  10^{30}{\rm cm}^{2}{\rm s}^{-1}$ in  quiescent milky-way-like galaxies (based on $\gamma$-ray observations and simulations; see \citealp{Lacki2011,Chan2019}; \hcr),
we see that  reasonable quasar luminosities will cause very effective CR confinement, with a strong ($\sim\! L^{1.4}$) dependence on luminosity up to 
the Bohm limit (at which point $\kappa_{\|} \sim 3\times 10^{24} {\rm cm}^{2}{\rm s}^{-1} B_{-8}^{-1}(R^{\rm CR}/1{\rm GV})$).

Such a strong increase in confinement may in turn have consequences for CR-driven winds or outflows. A crude measure of the
impact is to compare the force per unit mass on the gas due to the dust (i.e., radiation pressure) 
to that due to the CRs. The former is $\rho_{d} \ws/t_{s}\approx 6.7\times10^{-38 }{\rm g}{\rm cm}^{-2}{\rm s}^{-2}\,(\mu_{0}/0.01)L_{13}n_{-4}/(r_{100}^{2}\bar{\rho}_{d;0})$ assuming a supersonic dust-gas drift as above ($\ws/t_{s}$ is approximately independent of $\ad$ for $\ad\lesssim \lambda_{\rm rad}$); 
the latter is $\nabla P_{\rm CR}=e_{\rm CR}/(3 l_{\rm CR})\approx 1.7\times 10^{-38}{\rm g}{\rm cm}^{-2}{\rm s}^{-2}\,(e_{\rm CR}/0.01{\rm eV}\,{\rm cm}^{-3})(l_{\rm CR}/100{\rm kpc})^{-1}$. We see that even for relatively conservative values for $e_{\rm CR}$ and $l_{\rm CR}$ -- i.e., values similar to 
those obtained with $\kappa_{\|} \gtrsim 10^{30}{\rm cm}^{2}{\rm s}^{-1}$ in \hcr -- the CR force is not vastly smaller than the radiation-pressure force. Thus, even a modest increase in CR confinement  caused by dust-induced scattering would likely cause the force from CRs to overwhelm that due to radiation pressure, since 
$e_{\rm CR}$ or $1/l_{\rm CR}$ need only increase by a factor of several for this to happen. 
In other words, this indirect effect of radiation pressure as a feedback mechanism for galaxies -- driving small-scale waves that enhance CR confinement to the point where they drive outflows -- may 
be more efficient than the  direct driving of outflows  through radiation pressure and dust drag \citep{Murray2005}. Under appropriate conditions, this could provide a strong feedback mechanism
controlled by the luminosity of the quasar or galaxy.

\section{Conclusion}\label{sec:conclusion}

This paper has considered the influence of astrophysical dust grains on the confinement of galactic cosmic rays (CRs).
The cause of the interaction between these two seemingly unrelated components of the galactic ecosystem is 
the small-scale parallel Alfv\'en wave. Parallel Alfv\'en waves at sub-${\rm AU}$ scales
act to scatter and confine $\sim\!{\rm GeV}$ CRs through gyro-resonant interactions and the streaming instability. 
They also interact strongly -- also through gyro-resonant interactions -- with small charged dust grains, which act to damp
the waves if the dust is nearly stationary with respect to the gas, or cause instability if the dust streams super-Alfv\'enically.
This interaction implies a link between grain size (which determines the dust gyrofrequency through the dust mass and charge) 
and the CR rigidity: smaller grains, which are gyro-resonant with smaller wavelength AWs, interact with lower-rigidity CRs.

The two possibilities (wave damping or  instability from the dust) could have opposite effects on the astrophysics of cosmic rays.
In the case of nearly stationary dust, for which the magnitude of the wave damping can be estimated with reasonable fidelity 
 in well-ionized regions of the ISM,
dust could significantly \emph{enhance} CR transport (decrease confinement). 
Using reasonable assumptions about dust mass distributions and charging, we find damping rates from dust up to an order of magnitude larger 
than wave damping from the background turbulence \citep{Farmer2004,Zweibel2017} or nonlinear damping \citep{Cesarsky1981}.
Because wave damping directly determines  transport levels for self-confined CRs, this 
suggests that CR diffusion coefficients could be significantly 
larger than what would be expected without dust. 
The influence of dust may thus go some way towards explaining recent findings that standard models
of CR self confinement are too efficient to explain gamma-ray and in-situ CR grammage observations (\hcr; \citealp{Chan2019,Hopkins2020a}).

In the opposite case of dust-driven instability, the scattering of CRs by waves produced by dust has the potential
to significantly \emph{reduce} CR transport (enhance confinement), although
our estimates are far less reliable in this case due to significant astrophysical uncertainties.
The situation could occur in the circumgalactic medium around a luminous  galaxy, where radiation pressure would 
cause dust to stream
 outwards  with super-Alfv\'enic relative velocities, exciting small-scale Alfv\'en waves that scatter CRs.  
 Simple estimates suggest that the outwards force on the gas produced by the enhanced coupling to CRs
 through this interaction
could be much stronger than the direct force from the dust/radiation pressure itself  \citep{Murray2005}.

In either of the above scenarios, the  relative mass density of dust plays a key role in 
determining CR transport, which thus implies a dependence of CR confinement on metallicity.
In the first case of wave damping through dust, as most relevant to the ionized ISM away from high-luminousity sources, we predict  more efficient CR escape in high-metallicity environments, because AWs are more rapidly damped  at higher dust densities.  The second scenario, 
more relevant to halos around highly luminous galaxies, predicts the opposite, with stronger confinement at high metallicity. In this context, it is worth mentioning the particularly weak CR confinement in the SMC and LMC, as  inferred from $\gamma$-ray observations \citep{Lacki2011}. 
Since the SMC and LMC have a 
 low metallicity compared to the Milky Way, this  
does not fit with the first scenario of increased dust-induced damping, although given the wide array 
of other morphological and historical  differences, the lower CR confinement is likely unrelated 
to the detailed CR transport physics \citep{Chan2019}.
A metallicity dependence to CR transport could be a key ingredient in understanding the 
efficacy of CR feedback at high redshift.

\acknowledgements{Support for JS  was
provided by Rutherford Discovery Fellowship RDF-U001804 and Marsden Fund grant UOO1727, which are managed through the Royal Society Te Ap\=arangi. Support for PFH was provided by NSF Collaborative Research Grants 1715847 \& 1911233, NSF CAREER grant 1455342, and NASA grants 80NSSC18K0562 and JPL 1589742.  EQ was supported in part by NSF grant AST-1715070 and a Simons Investigator award from the Simons Foundation. 
We wish to acknowledge the hospitality of the Kavli Institute for Theoretical Physics during the program ``Multiscale Phenomena in Plasma Astrophysics,'' which was supported in part by the National Science Foundation under Grant No. NSF PHY-1748958.
}

\paragraph*{Data availability}No new data were generated or analysed in support of this research.

\appendix

\section{Growth and damping with a spectrum of grain sizes}\label{app:calc}

\begin{figure}
\begin{center}
\includegraphics[width=1.0\columnwidth]{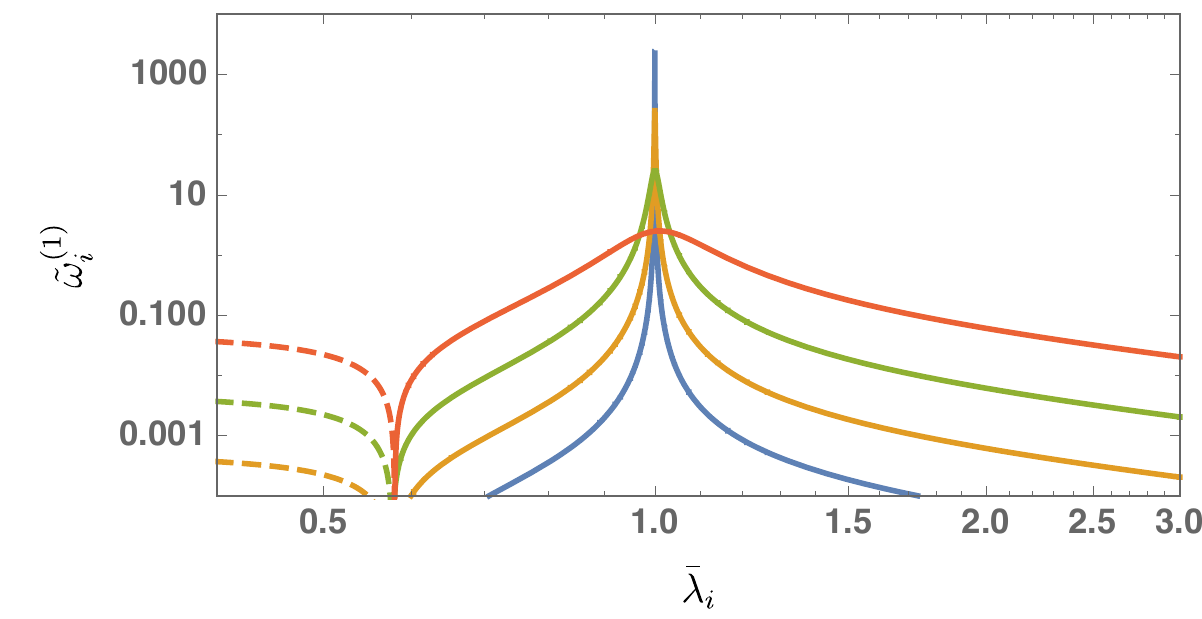}
\includegraphics[width=1.0\columnwidth]{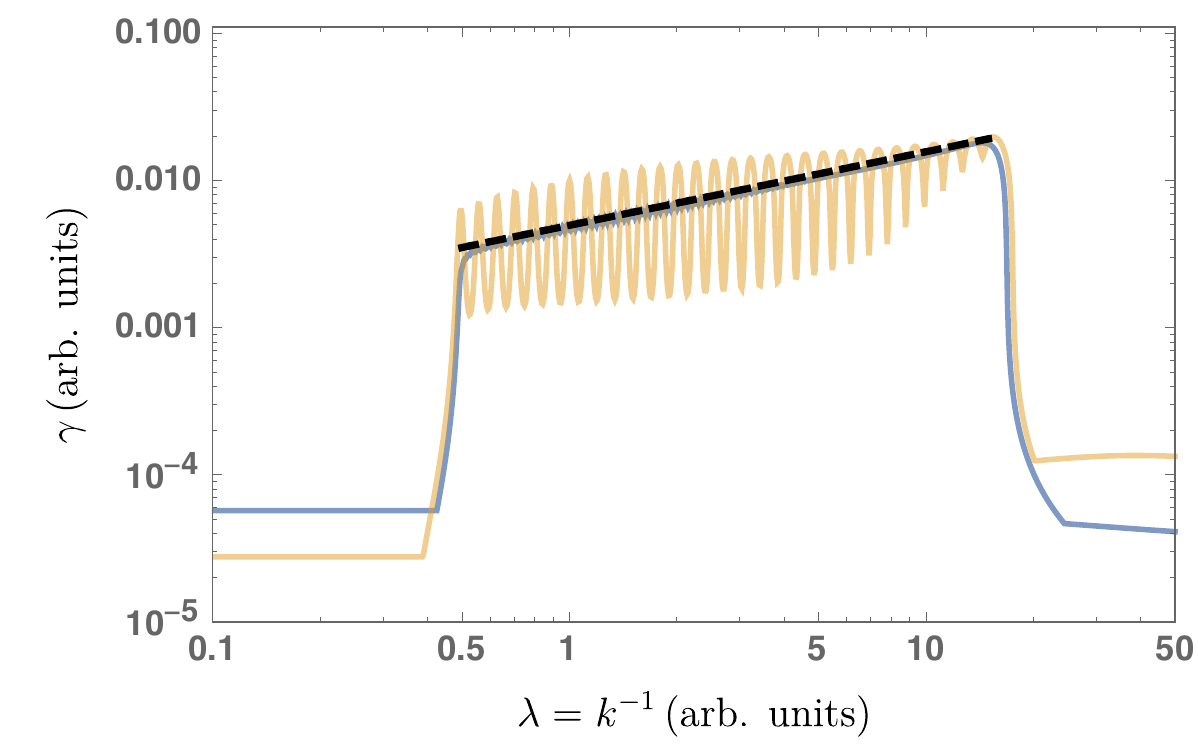}
\caption{Top panel: single-grain contribution to the growth rate $\tilde{\omega}^{(1)}_{i}$ (\cref{eq:discrete.growth.rate}) 
for $\machi=1$, with $\bar{t}_{s,i}=t_{s,i}\omega_{A}$ in the range $\bar{t}_{s,i}=\{10^{4},\,10^{3},10^{2},10^{1}\}$ (blue, orange, green, and red curves, respectively). Dashed lines show where $\tilde{\omega}^{(1)}_{i}<0$ ($\tilde{\omega}^{(1)}_{i}\approx \bar{t}_{s,i}^{-1}$ for $\bar{t}_{s,i}\gg1$). Bottom panel: comparison of the continuum approximation \cref{eq:growth.rate} (black dashed line) to the full dispersion 
relation computed numerically from the linearization of \crefrange{eq:MHD.dust.rho}{eq:MHD.dust.b}. The blue 
line shows the dispersion relation with 150 dust species, while the light-orange line shows the case with 30 dust species, illustrating the convergence towards the continuum solution. 
We use arbitrary units with $\ad$ ranging from $\ad=1$ to $\ad=30$ in logarithmically spaced increments, $v_{A}=1$, $\beta=10$,
$t_{L}=0.5 \ad^{0.5}$, $\mach=\ad^{0.5}$, $t_{s}=30 \ad$, a mass distribution $d\bar{\mu}/d\ln \ad\propto \ad^{0.5}$, and a total
dust density $\mu_{0}=0.01$. The analytic expression for the growth rate follows from inverting the resonance condition, $v_{A}t_{L}\mach=k^{-1}$, to give $\ad=2/k$, which gives $\gamma\approx 0.5\mu_{0}k^{-1/2}$ using \cref{eq:growth.rate}. }
\label{fig:app:growth.rate}
\end{center}
\end{figure}

In this appendix, we detail the derivations of \cref{eq:discrete.growth.rate,eq:growth.rate}, explore some of the 
properties of these, and discuss some  limits on their validity. We first discuss some technical features
of the discrete growth rate calculation in \cref{subapp: discrete}, then how to take the continuum limit in \cref{subapp: contin}. \revchng{The possible influence   of a dust velocity dispersion induced by turbulence is considered in \cref{app:subsub vel dispersion}.}
Although the general procedure -- computing a discrete growth rate first, then 
taking the limit $N_{d}\rightarrow\infty$ -- may seem more complex than starting directly from equations
for the continuum limit of  
\crefrange{eq:MHD.dust.rhod}{eq:MHD.dust.b}, the method seems to have some important advantages. Firstly,  in similar 
calculations  of dust interacting with an ion-electron plasma only through 
electromagnetic fields (i.e., without drag), but taking the limit $N_{d}\rightarrow\infty$ before computing a dispersion relation, \citet{Tripathi1996,Cramer2002} 
find  expressions for the growth rate that can be evaluated analytically only for  specific power-law dust-density distributions.
Second, our calculation provides useful physical intuition on its limits at finite 
dust stopping time $t_{s}$, explaining why the result is effectively independent of $t_{s}$ in the regimes of interest. 

\subsection{Discrete distribution}\label{subapp: discrete}
As discussed in \cref{sub: growth rate calc}, the growth rate of a discrete collection of grains can be straightforwardly 
obtained by computing the dispersion relation from the linearization of \crefrange{eq:MHD.dust.rhod}{eq:MHD.dust.b}.
Assuming that the  solution of the resulting polynomial is of the form $\omega=k v_{A}+ \omega^{(1)}$, where 
$\omega^{(1)}\sim \mathcal{O}(\mu)$, then expanding in $\mu\ll 1$, assuming all individual dust densities $\mu_{i}$ scale with $\mu$, leads to 
\begin{align}
\mu&\left( -\omega^{(1)}+ \sum_{i=1}^{N_{d}}\mu_{i}\omega_{A}\machi \frac{t_{L,i}^{2}(1+i \omega_{A}{t}_{s,i}\machi) - t_{s,i}^{2}\machi}{t_{s,i}^{2}+ t_{L,i}^{2}(1+i \omega_{A}{t}_{s,i}\machi)^{2}}\right)\nonumber \\
&\qquad\times \prod_{i=1}^{N_{d}} \machi (1+i\omega_{A}{t}_{s,i}\machi) + \mathcal{O}(\mu^{2})=0.\label{eq: DR in mu}
\end{align}
Solution of the first term in the product gives  \cref{eq:discrete.growth.rate}, after normalizing
the wavenumber of the mode to the resonant wavenumber using 
$\omega_{A}=k\,v_{A}=(t_{L}\machi\bar{\lambda}_{i})^{-1}$, where $\bar{\lambda}_{i}\lambda_{{\rm res},i}=k^{-1}$
and $\lambda_{{\rm res},i}\equiv v_{A}t_{L,i}\machi$.  

\revchng{\Cref{eq: DR in mu} is valid when  $\mu_{i}$ is sufficiently small  such that the effect of the dust gyromotion can be considered a small perturbation to the Alfv\'en wave. At larger  $\mu_{i}$, the structure of the linear system is modified because the matrix becomes defective (non-diagonalizable), causing the
growth rate to scale with $\sim \!\mu_{i}^{1/2}$ rather than the assumed $\sim \!\mu_{i}$, and thus invalidating the expression \eqref{eq: DR in mu}.}
As discussed in \citet{Squire2018} (see  discussion below their equation (12)), mathematically this implies that the growth rate of the mode must be less than that of the ``resonant 
drag instability'' between the dust gyromotion and the Alfv\'en wave, which has the growth rate, 
 \begin{equation}
\Im(\omega) \approx \frac{1}{2}\mu_{i}^{1/2}|t_{L,i}|^{-1}|\machi|^{1/2}=\frac{1}{2}\mu_{i}^{1/2}|\machi|^{3/2} \omega_{A}|_{k=\lambda_{{\rm res},i}^{-1}}\label{eq: RDI growth rate}
\end{equation}
at resonance ($k=\lambda_{{\rm res},i}^{-1}$)
for a single  grain population $i$ (\hsmag). In other words,  \cref{eq: RDI growth rate}, rather than \cref{eq:discrete.growth.rate}, is the correct expression at larger $\mu_{i}$.  Equations \eqref{eq: DR in mu} and \eqref{eq:discrete.growth.rate}
are thus valid if 
$\tilde{\omega}_{i}^{(1)}|_{\bar{\lambda}_{i}=1}\lesssim \mu_{i}^{-1/2} \machi^{3/2}$,
or -- expanding $\tilde{\omega}_{i}^{(1)}$ in $\bar{t}_{s,i}\gg1$ to find $\tilde{\omega}_{i}^{(1)}|_{\bar{\lambda}_{i}=1}\approx \machi^{3}\bar{t}_{s,i}/2+\mathcal{O}(\bar{t}_{s,i}^{-1})$ -- for 
\begin{equation}
|\machi|^{3/2}\bar{t}_{s,i}\lesssim \mu_{i}^{-1/2}.\label{eq: condition on growth rate}
\end{equation}
This condition compares well against full numerical  solutions to the dispersion relation.

\subsection{Continuum limit}\label{subapp: contin}

For $\bar{t}_{s,i}\gg1$, the contribution to growth or damping from a single grain species (the function $\tilde{\omega}^{(1)}_{i}$ in the sum of  \cref{eq:discrete.growth.rate}),  becomes very sharply 
peaked around $\bar{\lambda}_{i}=1$. For illustration, we plot $\tilde{\omega}^{(1)}_{i}(\bar{\lambda}_{i})$ 
in the top panel of  \cref{fig:app:growth.rate}, showing (i) how its peak value (at $\bar{\lambda}_{i}=1$) scales linearly with $\bar{t}_{s,i}$ [$\tilde{\omega}^{(1)}_{i}(1)\approx \machi^{3}\bar{t}_{s,i}/2+\mathcal{O}(\bar{t}_{s,i}^{-1})$], (ii) that $\tilde{\omega}^{(1)}_{i}\rightarrow 0$ as $\bar{\lambda}_{i}\rightarrow \infty$, and (iii) that $\tilde{\omega}^{(1)}_{i}(0)\approx -\bar{t}_{s,i}^{-1}+\mathcal{O}(\bar{t}_{s,i}^{-3})$. Moreover, it transpires
that 
\begin{equation}
\int_{0}^{\infty}d\bar{\lambda}_{i}\tilde{\omega}^{(1)}_{i}=\frac{\pi}{2} \sgn(\machi)\machi^{2},
\end{equation}
implying that at $\bar{t}_{s,i}\gg1$, $\tilde{\omega}^{(1)}_{i}$ is well approximated by a delta function, with no $\bar{t}_{s,i}$ dependence: $\tilde{\omega}^{(1)}_{i}\approx (\pi/2) \sgn(\machi)\machi^{2}\delta(\bar{\lambda}_{i}-1)=  (\pi/2) \sgn(\machi)\machi^{2}\lambda_{{\rm res},i}\delta(k^{-1}-\lambda_{{\rm res},i})$.  
Given that 
the total growth rate from $N_{d}$ species is simply the sum of the individual growth rates, we can 
derive an expression for  the continuum damping/growth rate, $\gamma(k)$, by converting the sum in  \cref{eq:discrete.growth.rate}
to an integral.
This is done by stipulating that
the area under  $\gamma(k)$ between resonances, $\gamma(k)d \lambda_{{\rm res},i} =\gamma(k)(\lambda_{{\rm res},i+1}-\lambda_{{\rm res},i})$ is equal to the area under the discrete growth rate at the same resonance, which is $\Im[\omega^{(1)}(\lambda_{{\rm res},i})]\,d\lambda_{{\rm res},i} \approx \mu_{i} (\pi/2) \omega_{A}\sgn(\machi)\machi^{2}\lambda_{{\rm res},i}$, 
where in the final step we have used the delta-function approximation for $\tilde{\omega}^{(1)}_{i}$.
Replacing $\mu_{i}$ by $d\mu$ with $\mu_{0}=\sum_{i=1}^{N_{d}}\mu_{i}=\int\,d\mu$, we get 
\begin{align}
\gamma(k) &= \frac{\Im[\omega^{(1)}(\lambda_{{\rm res},i})]\,d\lambda_{{\rm res},i}}{d \lambda_{{\rm res},i}} \nonumber \\
&= \omega_{A}\frac{\pi}{2}\sgn(\mach)\frac{d\mu}{d\ln \ad}\left( \frac{d \lambda_{\rm res}}{d\ln \ad}\right)^{-1}  \mach^{2}(\ad)\lambda_{{\rm res}}(\ad)\nonumber \\
&=\omega_{A}\frac{\pi}{2} \sgn(\mach)\frac{d\mu}{d\ln \ad} \left(\frac{d\ln |t_{L}|}{d\ln \ad} +\frac{d\ln |\mach|}{d\ln \ad} \right)^{-1}\mach^{2} ,\label{eq: growth rate derivation}
\end{align}
where all functions of $a$ ($t_{L}(\ad)$, $\mach(\ad)$ etc.) are converted to functions of $k$ by inverting $k^{-1}=\lambda_{\rm res}(\ad) = v_{A}t_{L}\mach$
(note that $k^{-1}$ can be negative, which allows for different mode polarizations depending on the sign of $t_{L}$ and $\mach$).

To understand possible restrictions on   \cref{eq: growth rate derivation}, let us  note
two key points that are implicit in the derivation. First, as we approach the true continuum limit, the 
condition \cref{eq: condition on growth rate} must become satisfied at some point, since we are taking 
the limit as $\mu_{i}\rightarrow 0$. Second, although we used the delta function approximation to $\tilde{\omega}^{(1)}_{i}$
to derive   \cref{eq: growth rate derivation}, it is clear that the final answer will be identical 
if the resonances overlap (finite $\bar{t}_{s}$): each of the nearby resonances from different sized grains
will contribute a small amount to the region  $d \lambda_{{\rm res},i}$, whose total -- assuming that the peak of $\tilde{\omega}^{(1)}_{i}$
does not vary much across the width of the contribution from a single grain -- will add up to the same 
as the delta function approximation. We thus see that the effect of finite $\bar{t}_{s}$ is to smooth out any 
sharp features in the continuum growth rate expression \cref{eq:growth.rate} or \cref{eq: growth rate derivation} by the width of an individual resonance, which 
is $d\bar{\lambda}_{i}\sim \bar{t}_{s,i}^{-1}$ or\begin{equation}
dk\sim \frac{k^{2}\lambda_{\rm res}}{\omega_{A}t_{s}}=k\,|\mach| \frac{|t_{L}|}{t_{s}}.
\end{equation}
In addition, minor (order $\bar{t}_{s}^{-1}$) inaccuracies arise due to $\tilde{\omega}^{(1)}_{i}$ becoming negative as $k\rightarrow\infty$.

We compare the continuum approximation \cref{eq:growth.rate} to numerical calculations of the full dust-gas dispersion 
relation in the bottom panel of  \cref{fig:app:growth.rate}. Parameters $t_{s}(\ad)$, $t_{L}(\ad)$, and $\mach(\ad)$ are chosen as arbitrary functions for illustrative purposes.
The comparison of the orange and blue curves, which are
the numerical calculations with 30 and 150 dust species respectively, 
illustrates the accuracy of the continuum approximation (black dashed line) in the region of resonances
once the individual resonances overlap (blue line). The continuum approximation predicts no growth 
or damping for wavenumbers outside of the band of resonances, which we see is not completely accurate (although growth rates are 
$\sim\! 10^{-2}$ lower outside the band).

\subsection{Dust velocity dispersion}\label{app:subsub vel dispersion}

\revchng{It is expected that even if $\ws=0$, the dust should have some non-zero velocity dispersion due to the effects
of turbulent gas motions \citep{Lazarian2002,Lee2017}, likely with the dispersion of perpendicular velocities 
dominating over that of parallel velocities for most magnetized turbulence  \citep{Yan2004}. Including this formally and in detail
requires recourse to a a fundamentally kinetic description the dust; this would involve linearizing a dust Vlasov equation
for its distribution function $f_{d}(\bm{x},\bm{v},\ad)$, which is beyond the scope of this work. 
However, in order to get a basic idea of whether such effects are likely to be important, we can consider
adding in a dust pressure term into the dust velocity evolution equation \eqref{eq:MHD.dust.v}. This must
have the form $\nabla_{d\perp}P_{d\perp,i} + \partial_{z}P_{d\|,i}\hat{\bm{z}}$, where $\nabla_{\perp}$ is the gradient in the perpendicular ($x$ and $y$) direction 
and $P_{d\perp,i}$ and $P_{d\|,i}$ refer to the dust velocity dispersions perpendicular and parallel to the field, respectively (the equilibrium off-diagonal components of the
dispersion tensor must be small if the dust velocity distribution is symmetric around the magnetic field).
While $P_{d\perp,i}$ and $P_{d\|,i}$ are not determined without recourse back to the kinetic description, a simple
isothermal closure -- $P_{d\perp,i}=\rho_{d,i}c_{d\perp0,i}^{2}$ and $P_{d\|,i}=\rho_{d,i}c_{d\|0,i}^{2}$, where $c_{d\perp0,i}$ and $c_{d\|0,i}$ are fixed -- 
should at least give an idea of whether the velocity dispersion should fundamentally change the modes of interest. 
Repeating the calculation above, one finds that it does not: the perpendicular pressure has no effect whatsoever, because
it depends only on perpendicular gradients and we explicitly focus only on parallel modes; the parallel pressure modifies  
the final part  of \cref{eq: DR in mu}, changing the term in the product to $\machi(1+i\omega_{A} t_{s,i}\machi)-i k_{z}^{2}t_{s,i}c_{d\|0,i}^{2}/\omega_{A}$,
but this has no effect on the modes of interest either (at small $\mu_{i}$).   
This illustrates that the direct effect of dust pressure is unimportant for the Alfv\'enic modes of interest, 
although it does not capture more complex kinetic effects such as the broadening 
of the distribution of resonances due to the range of parallel velocities. The mode's indifference to 
the perpendicular velocity distribution is straightforward to understand physically: the dust's gyro-time, 
which determines the damping through the resonance with the Alfv\'en wave ($\bar{\lambda}_{i}=1$), 
is independent of its perpendicular velocity. 
}

\bibliographystyle{apj} 
\bibliography{fullbib_formatted}


\end{document}